\documentclass[fleqn,usenatbib]{mnras}

\usepackage{newtxtext,newtxmath}
\usepackage[T1]{fontenc}
\usepackage{CJK}
\usepackage{graphicx}
\usepackage{amsmath}
\usepackage{xcolor}
\usepackage{graphicx}
\usepackage{caption}
\usepackage{subcaption}

\definecolor{darkgreen}{RGB}{0,127,0}

\title[Eccentricity in MHD simulation of superhump system]{Investigating lack of accretion disk eccentricity growth in a global 3D MHD simulation of a superhump system}
\author[B. Oyang, Y.-F. Jiang, and O. Blaes]{
Bryance Oyang,$^{1}$\thanks{E-mail: b\_oyang@ucsb.edu}
Yan-Fei Jiang(\begin{CJK*}{UTF8}{gbsn}姜燕飞\end{CJK*})$^{2}$
and Omer Blaes$^{1}$
\\
$^{1}$Department of Physics, University of California, Santa Barbara, CA 93106, USA\\
$^{2}$Center for Computational Astrophysics, Flatiron Institute, New York, NY 10010, USA
}

\date{}
\pubyear{2020}

\begin{document}

\label{firstpage}
\pagerange{\pageref{firstpage}--\pageref{lastpage}}
\maketitle

\begin{abstract}
    We present the results of a 3D global magnetohydrodynamic (MHD) simulation of an AM CVn system that was aimed at exploring eccentricity growth in the accretion disc self-consistently from a first principles treatment of the MHD turbulence. No significant eccentricity growth occurs in the simulation. In order to investigate the reasons why, we ran 2D alpha disc simulations with alpha values of 0.01, 0.1, and 0.2, and found that only the latter two exhibit significant eccentricity growth. We present an equation expressing global eccentricity evolution in terms of contributing forces and use it to analyze the simulations. As expected, we find that the dominant term contributing to the growth of eccentricity is the tidal gravity of the companion star. In the 2D simulations, the alpha viscosity directly contributes to eccentricity growth. In contrast, the overall magnetic forces in the 3D simulation damp eccentricity.
    We also analyzed the mode-coupling mechanism of Lubow, and confirmed that the spiral wave excited by the 3:1 resonance was the dominant contributor to eccentricity growth in the 2D $\alpha=0.1$ simulations, but other waves also contribute significantly. We found that the $\alpha=0.1$ and 0.2 simulations had more relative mass at larger radii compared to the $\alpha=0.01$ and 3D MHD simulation, which also had an effective $\alpha$ of 0.01. This suggests that in 3D MHD simulations without sufficient poloidal magnetic flux, MRI turbulence does not saturate at a high enough $\alpha$ to spread the disc to large enough radii to reproduce the superhumps observed in real systems.
\end{abstract}


\section{Introduction}
Cataclysmic variable (CV) systems have long been a major testing ground for accretion disc theory.  Fundamental to this theory is the mechanism of angular
momentum transport, with magnetorotational (MRI, \citealt{mri}) turbulence, spiral waves excited by the tidal field of the companion star \citep{lin79}, and possibly magnetized outflows (e.g. \citealt{sce20}) all playing roles.  The time scales of dwarf nova outbursts provide the strongest constraints on the \citet{alpha_disk} $\alpha$-parameterization of angular
momentum transport \citep{KIN07,KOT12}, with significantly higher values of $\alpha$ in outburst than in quiescence.  CVs, like many accretion disc systems, also exhibit broadband aperiodic variability with a linear rms-flux relation \citep{sim12}, indicative of radially propagating stochastic fluctuations in accretion rate \citep{lyu97}.  Periodicities are also commonly observed in some systems that shed light directly on the dynamics of the disc, the strongest being so-called superhumps (e.g. \citealt{patterson}): oscillatory variations in brightness with periods slightly longer than the binary orbital period.  Negative superhumps with periods slightly shorter than the orbital period are also observed (e.g. \citealt{sma09b}).


Early simulation and theoretical work showed that (positive) superhumps are likely due to an eccentric disc undergoing apsidal precession, with the superhump frequency being related to the orbital and apsidal precession frequencies by $\Omega_\text{superhump} = \Omega_\text{binary} - \omega_\text{precession}$ \citep{osaki, whitehurst}. The tidal gravity of the donor star can excite eccentricity of the disc through a mode coupling between spiral waves excited at the 3:1 resonance in the disc and the tidal potential \citep{lubow_theory}.  \citet{bis04,kai06} propose an alternative explanation in terms of an eccentric wave, and \citet{sma09a,sma10} suggests that superhumps are caused by the accretion stream interacting with a disk with nonaxisymmetric vertical thickness.  Both alternatives do not require the existence of the 3:1 resonance, but they still ultimately rest on an association with apsidal precession of the disc.

Given this dynamical association, the superhump periods have been used to estimate the mass ratios $q$ of the binary systems \citep{patterson, kato_eps_to_q}, but these estimates might neglect important differences between systems such as effective viscosity, pressure forces, and mass distribution within the disc which can influence the apsidal precession rate \citep{mur98, goodchild_ogilvie}. Many details of superhumps remain inadequately explored in simulations, such as their period changes during the course of an outburst \citep{kato_survey}. However, smoothed particle viscous hydrodynamics simulations have been successful in producing variations of the global dissipation in the disc that resemble lightcurves of superhumps \citep{mur96,mur98,sim98,smi07}.

\citet{kley} performed 2D grid-based fluid simulations with an explicit kinematic viscosity exploring eccentricity evolution in relation to superhumps.  They found that viscosity plays an important role, with larger viscosity resulting in more rapid eccentricity growth. Given that the true nature of viscosity is likely MRI turbulence, it would be interesting to see whether a global 3D MHD simulation treating the angular momentum transport from first principles can replicate the observed superhump lightcurves and give insights into the superhump phenomenon. Of additional interest is evidence that eccentric waves can themselves interact with and dampen MRI turbulence \citep{dewberry}.

Several recent advances have been made in simulating accretion discs in CVs with MHD. \citet{ju_stone1,ju_stone2} explored the relative importance of spiral shocks and MRI turbulence in driving accretion in vertically unstratified global MHD simulations of CVs, and \citet{patryk} extended this work to include vertical stratification.
Significant numerical challenges remain, however, in resolving the scale height for realistic disc temperatures, achieving realistic Prandtl numbers, and running the simulations for long enough to achieve inflow equilibrium.
These difficulties result in unavoidable idealizations and approximations. While simulations can inform us about the detailed behavior of these systems and provide explanations for observed phenomena, observations can conversely be used to constrain simulations and ensure that approximations made in the simulations do not result in excessive deviation from reality.

 AM Canum Venaticorum (AM CVn) stars are very compact binary star systems with a short (typically $\lesssim 1$ hour) orbital period in which a white dwarf primary accretes gas from a helium rich secondary donor star \citep{solheim}. The small spatial dynamic range and short dynamical timescales of AM CVns makes them particularly attractive targets for global numerical simulations seeking to understand the nonlinear physics in accretion discs.  Observationally, the shortest period, persistent high state AM CVns exhibit superhumps \citep{solheim}. 
Motivated by these considerations, we performed a global 3D MHD simulation of an AM CVn accretion disc modeled after the system SDSS J1908 \citep{fontaine}, which has a binary mass ratio of $q = 0.1$ and exhibits permanent superhumps.  We therefore expected our simulated disc to also develop an eccentric, slowly precessing disc. However, despite continuing the simulation for over 200 binary orbits, our disc remains mostly circular, with no obvious sign of eccentricity growth over time. The lack of eccentricity growth in MRI simulations of CV discs has also been previously reported in \citet{ju_stone2}, though there they used a larger binary mass ratio of $q = 0.3$.
The primary goal of this paper is then to better understand why the 3D MHD simulations do not produce eccentricity growth and to provide tools for future simulations wishing to resolve this discrepancy.

This paper is organized as follows: Section \ref{sec:method} describes our setup and numerical methods. Section \ref{sec:general_3d} describes general properties of our 3D MHD disc to enable comparisons with similar simulations. Section \ref{sec:eccent} aims to explore in detail the eccentricity growth mechanism and drivers through a comparison of the 3D MHD simulation with 2D alpha disc simulations that do exhibit growing eccentricity. Analysis of the mode-coupling mechanism of \citet{lubow_theory} is also made. Section \ref{sec:conclusion} summarizes our main results and suggests future directions.


\section{Method}\label{sec:method}
Our 3D MHD simulation is done using the code Athena++ \citep{athena} using a spherical polar grid. We use $r, \theta, \phi$ to denote the radial, polar angle, and azimuthal angle coordinates respectively. The advantage of using polar coordinates is that smaller radii remain resolved, and Athena++ is designed to conserve the $z$-component of angular momentum in polar coordinates \citep{athena}, aiding in the long-timescale numerical accuracy. We adopt a non-inertial rotating frame of reference with the origin at the center of the primary white dwarf. In these coordinates, the binary companion and tidal potential are stationary in time.

Because previous numerical work has found that eccentricity grows in 2D simulations of binary systems, we also run our own 2D $\alpha$ viscosity disc simulations in cylindrical polar coordinates to act as a basis of comparison for the 3D MHD simulation. The 2D simulations are a tool to better understand why eccentricity grows in these binary systems, and therefore also to understand why eccentricity does not grow in the 3D MHD simulation.

\subsection{Units and scaling}

\begin{table}
\centering
\caption{Simulation units to cgs conversion: multiply by these to get cgs quantities}
\label{tab:unit_conv}
\begin{tabular}{ll}
\hline
Quantity & Conversion\\
\hline
Distance & $4.69 \times 10^{8}$ cm\\
Density & $1.00\times 10^{-4}$ g cm$^{-3}$\\
Temperature & $5.00 \times 10^4$ K\\
Pressure & $3.05 \times 10^8$ dyn cm$^{-2}$\\
Velocity & $1.75 \times 10^6$ cm s$^{-1}$\\
B field & $6.19 \times 10^{4}$ G\\
\hline
\end{tabular}
\end{table}

Though our current MHD simulations have not yet approached the temperature regime of real AM CVn systems, an eventual goal of these simulations will be to include realistic thermodynamics and radiative transport for direct comparison with observations. To facilitate this goal and to enable comparisons with future simulations, Table~\ref{tab:unit_conv} summarizes the units used in our simulations here.

We choose our distance unit to be the white dwarf radius, which we take to be $4.69\times10^8$~cm. Our density unit choice together with distance gives a mass unit. The temperature unit is chosen roughly according to the effective temperature of the discs as seen in observations. The pressure unit is chosen such that $k_B / \mu m_p = 1$ so that $P = \rho T$. The velocity and hence time unit is chosen as the sound speed at one unit of temperature. The magnetic field unit is chosen such that $B^2 / 2$ in code units gives the magnetic pressure in code units.

In the remainder of this paper, numbers without explicit units are given in these simulation units.

\subsection{Binary system parameters}
We model our binary after the AM CVn discovered in \citet{fontaine}. We let the primary mass be $1.1$ M$_\odot$ and the secondary mass be $0.11$ M$_\odot$ for a binary mass ratio of $q = 0.1$. We set the binary period to be as measured in \citet{fontaine} of $938.5$ seconds (though \citet{kupfer} determined the period spectroscopically and found it to be slightly larger at 1085.7~s).
In code units, we use a binary separation of $a = 32.68$ and assume the binary has zero eccentricity. These give an $L_1$ Lagrange point at a distance of $23.45$ from the white dwarf and a nominal 3:1 resonance at a distance of $15.2$. The $L_1$ point is the location where the gas from the secondary spills over to the primary to form the accretion disc, and a sphere at that radius is used as the outer boundary of the simulation domain.

\subsection{Equations solved}
For the 3D model, we solve the ideal MHD equations for a locally isothermal gas in the rotating frame.
\begin{subequations}
    \begin{gather}
        \partial_t \rho + \nabla \cdot (\rho \mathbf{v}) = 0\\
        \partial_t (\rho \mathbf{v})
        + \nabla \cdot (\rho \mathbf{v v} - \mathbf{BB} + P^*\mathbf{I})
        = -2 \rho \Omega_p \mathbf{z} \times \mathbf{v}
        - \rho \nabla \Phi\\
        \frac{\partial \mathbf{B}}{\partial t} - \nabla \times (\mathbf{v} \times \mathbf{B}) = 0\\
        P^* = \rho T + \frac{\mathbf{B}^2}{2}\\
        T = \frac{GM_1}{R} \left( \frac{H}{R} \right)^2 \propto \frac{1}{R}
    \end{gather}
The temperature at each radius was chosen such that the disc scale height $H/R=0.044$ is constant. The potential is given by
\begin{align}
    \Phi &= -\frac{GM_1}{r} - \frac{GM_2}{|\mathbf{r} - \mathbf{R_2}|} - \frac{1}{2} (\Omega_p r \sin\theta)^2
    + \frac{GM_2}{R_2^3} (\mathbf{R_2} \cdot \mathbf{r})
\end{align}
\end{subequations}
where $\mathbf{R_2}=32.68 \mathbf{\hat{x}}$ is the location of the secondary and $\Omega_p$ is the binary angular velocity.

For the 2D model, we solve the vertically integrated versions of the 3D equations with the potential restricted to the midplane. To allow for a parameter exploration of the angular momentum transport mechanism in 2D, instead of using magnetic fields, we replace the Maxwell stress tensor with an $\alpha$ viscosity represented by a viscous stress tensor $\mathbf{\Pi}$ whose components in a Cartesian basis are given as
\begin{subequations}
\begin{gather}
    \Pi_{ij} = \rho \nu \left(\frac{\partial v_i}{\partial x_j}
    + \frac{\partial v_j}{\partial x_i}
    - \frac{2}{3} \delta_{ij} \nabla \cdot \mathbf{v} \right)
\end{gather}
The kinematic viscosity $\nu$ has a radial dependence provided by a standard $\alpha$ prescription.
\begin{gather}
    \nu(R) = \alpha T \sqrt{\frac{R^3}{GM_1}}
\end{gather}
\end{subequations}
We used three values of $\alpha=0.01, 0.1, 0.2$.

The equations are solved using Athena++'s HLLC Riemann solver with second-order spatial reconstruction and the second-order van Leer time integrator. The gas internal energy is reset at the end of each timestep to enforce the locally isothermal condition. A density floor of $\rho_\text{floor} = 10^{-8}$ and a gas pressure floor of $P_\text{floor} = 10^{-10}$ were chosen for both the 2D and 3D simulations.

\subsection{Initial and boundary conditions}

\subsubsection{3D setup}
The 3D simulation used a spherical polar grid that spanned $(r, \theta, \phi) \in [1, 23.4] \times [0, \pi/2] \times [0, 2\pi)$. The root computational domain was subdivided into $64 \times 64 \times 128$ cells. Two levels of static mesh refinement, resulting in $2^2 = 4$ times the resolution, were used between $\theta \in [0.7, 2.4]$. A logarithmically spaced radial grid was used to maintain cells' aspect ratio and to better resolve the regions near the primary white dwarf.

For our 3D simulation, we wished to build up all the mass in the accretion disc from the secondary's gas inflow stream. We inject gas from the binary's $L_1$ point as a radial boundary condition. Based on the analytic work of \citet{lubow_shu_stream}, a gaussian density profile was set in the ghost cells centered on the $L_1$ point and the gas is given an inward radial velocity. Away from the $L_1$ point, we copy the density, azimuthal velocity, and pressure from the last cell in the computational domain into the ghost cells. In order to minimize inflow of mass from the boundary, the radial velocity in the ghost cells is copied from the last cell in the computational domain only if fluid is moving out of the simulation domain but is set to $0$ otherwise. The same is done for the inner radial boundary.

Magnetic field loops were also initially injected from the $L_1$ point. The magnetic field in the ghost zones of the outer radial boundary is initialized with a vector potential proportional to the density in the stream. The amplitude is determined to make sure magnetic pressure in the middle plane of the stream is 5\% of the gas pressure. Outside of the stream, radial magnetic field is copied from the last active zone to the ghost zones while both the poloidal and azimuthal components are set to be 0. For the inner radial boundary, we also copy the radial magnetic field component from the last active zone to the ghost zones and set all the other components in the ghost zones to be 0.

More details of the changes made to the 3D simulation over the course of the simulation are explained in Section~\ref{sec:general_3d}.

\subsubsection{2D setup}\label{sec:2d_setup}
The 2D simulation used a cylindrical polar grid that spanned $(r, \phi) \in [1, 23.4] \times [0, 2\pi)$. The computational domain was subdivided into 256$ \times $512 cells, but no additional mesh refinement is used in 2D. As with the 3D simulation, a logarithmically spaced radial grid and uniformly spaced azimuthal grid was used.

The 2D simulation was initialized based on data from the 3D simulation to
enable comparisons of the two. All quantities in the 2D simulation were
initialized with symmetry in $\phi$. We use the 3D quantities at simulation
time $t = 226$ binary orbits. The initial 2D surface density $\rho_\text{2D}$ was set from the
3D volume density $\rho_\text{3D}$ as
\begin{gather}
    \rho_\text{2D}(r) = \frac{1}{2\pi} \int_0^{2\pi} d\phi \int_0^\pi d\theta\, \rho_\text{3D}(r, \theta, \phi) r \sin\theta
\end{gather}
The vertical integration was done in $\theta$ rather than $z$ since the 3D
simulation data naturally lies in a spherical polar grid. This
approximation was deemed sufficient since most of the mass is concentrated
near the midplane in the 3D simulation.

The initial radial velocity was set to $0$. Both the 2D azimuthal velocity and
temperature were set according to the midplane values of the 3D simulation, azimuthally
averaging the former quantity.

In the 2D simulations, we do not use a gas inflow stream. At the radial boundaries, we copy the density, azimuthal velocity, and pressure from the last cell in the computational domain into the ghost cells. In order to minimize inflow of mass from the boundary, the radial velocity in the ghost cells is copied from the last cell in the computational domain only if fluid is moving out of the simulation domain but is set to $0$ otherwise.

\section{3D MHD simulation: general properties}\label{sec:general_3d}
Though the main focus of this paper is to explore eccentricity growth in numerical simulations of binary systems, we give a general description of our 3D MHD simulation here.
To enhance the Maxwell stress, we have also tried to add additional magnetic field loops in the simulation domain directly. When we restart the simulation, we add additional magnetic fields based on 
a vector potential, which is only non-zero for $8<r<12$. The toroidal component of the vector potential is set to be proportional to density and the amplitude is chosen so that the magnetic pressure of the 
new magnetic field component in the disk midplane is about 10 percent of the gas pressure. Notice that we do not change the existing magnetic field in the simulation and final magnetic field after the change 
can still maintain the $\nabla \cdot B = 0$ condition numerically.

\begin{figure}
    \includegraphics[width=\columnwidth]{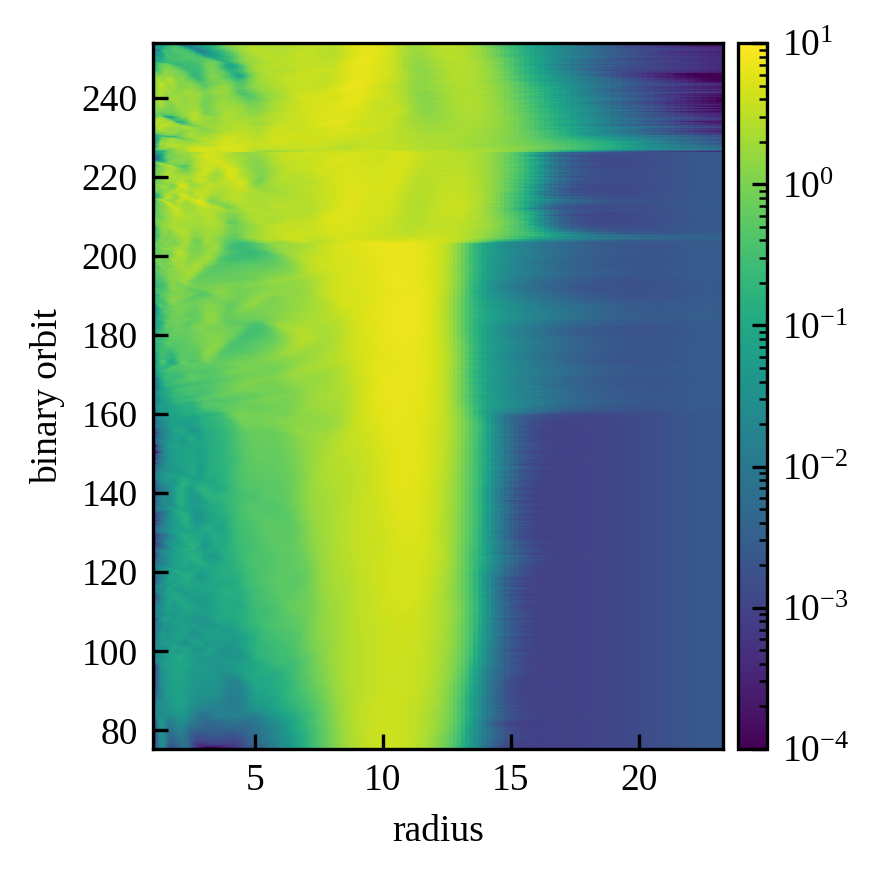}
    \caption{3D MHD vertically integrated density in radius and time. Times where artificial changes are manually made to the magnetic field to help the disc spread are clearly visible as sharp horizontal transitions. At $t=226$ binary orbits, the accretion stream is also turned off to make it easier for eccentricity to grow.}
    \label{fig:spacetime_surf_dens}
\end{figure}

Figure~\ref{fig:spacetime_surf_dens} shows a spacetime plot of the vertically integrated density in the second half of the simulation. The disc mass is accumulated over time from the accretion stream injected at the $L_1$ outer boundary condition.

\begin{figure}
    \includegraphics[width=\columnwidth]{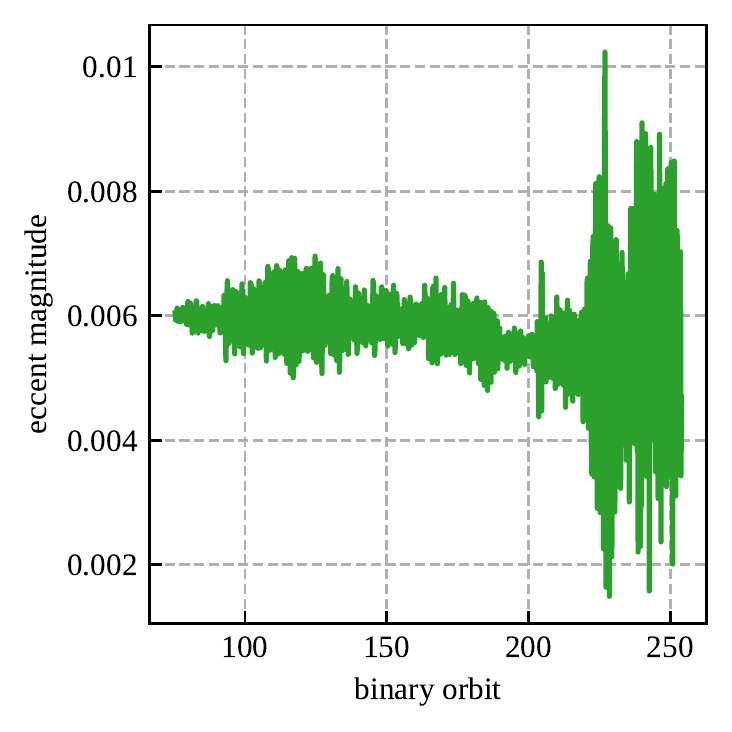}
    \caption{3D MHD globally mass-averaged eccentricity over time. At $t=226$ binary orbits, the accretion stream is turned off and magnetic field is manually added to make it easier for eccentricity to grow, but eccentricity remains small.}
    \label{fig:mhd_3d_eccent_history}
\end{figure}

In the prior 2D grid-based simulations in \citet{kley}, eccentricity in their similar binary system tended to saturate approximately around $\sim 0.1$ after a time on the order of a few hundred binary orbits, depending on their choice of  kinematic viscosity, disc temperature, boundary conditions, and binary mass ratio. But as seen in Figure \ref{fig:mhd_3d_eccent_history}, our 3D MHD disc displays no significant global eccentricity at any point during the simulation.

Since previous results such as \citet{lubow_stream_impact} and \citet{kley} found that a constant gas inflow stream from the binary's $L_1$ point damps eccentricity, and \citet{kley} found that increasing viscosity increases the growth rate of eccentricity, we turned off the stream after around $t=226$ binary orbits and also added additional magnetic field loops into the disc to determine if these could help induce eccentricity growth in the 3D MHD simulation. However, still no significant eccentricity develops, and this is discussed in detail in Section~\ref{sec:eccent}.

In the remainder of this section, we give a brief description of the state of our 3D MHD simulation at $t=225$ binary orbits right before the stream is turned off to enable comparisons with past and future binary disc MHD simulations. In particular, \citet{patryk} have run 3D MHD simulations of a CV also using Athena++, though with a mass ratio of $q = 0.3$, lower Mach numbers of 5 and 10 at the inner boundary compared to our Mach number of 20, and other different numerical details.  

\begin{figure}
    \includegraphics[width=\columnwidth]{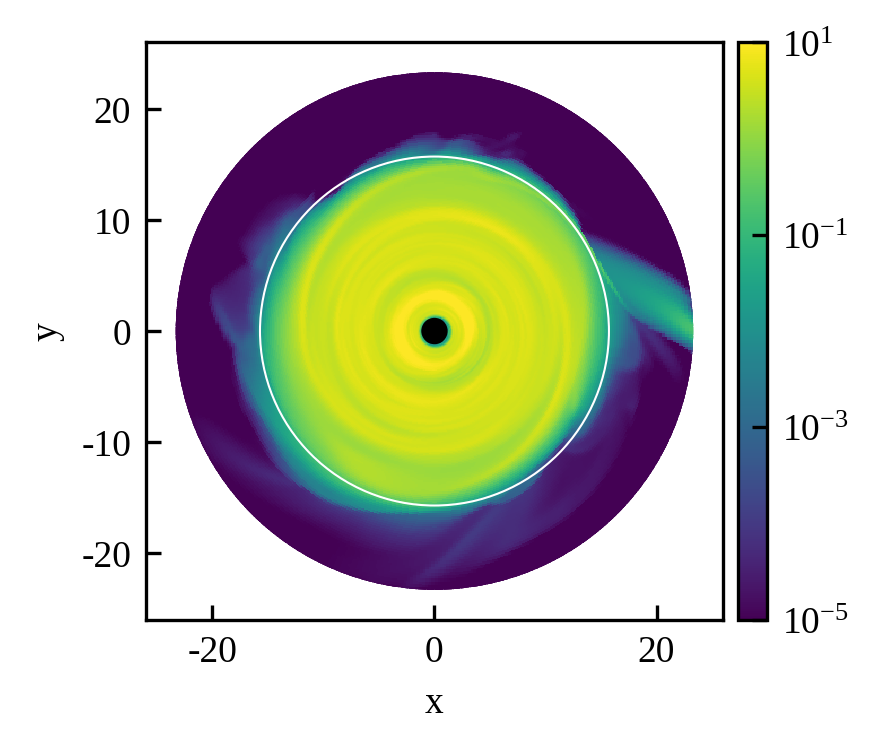}
    \caption{3D MHD midplane volume density at $t = 225$ binary orbits. The secondary 
    donor star is located at $(x, y) = (32.68, 0)$ (outside the simulation domain). The nominal 3:1 resonance is located at $r = 15.2$, indicated by white circle.}
    \label{fig:mhd_3d_cone_dens}
\end{figure}

We plot the midplane density of the 3D MHD simulation in Figure~\ref{fig:mhd_3d_cone_dens}. An eccentric disc can be described as an off-centered slowly precessing ellipse pattern in the non-rotating frame, but this is not seen here. Detailed measurements of the eccentricity growth over time are given in Section~\ref{sec:eccent}.

\begin{figure}
    \includegraphics[width=0.95\columnwidth]{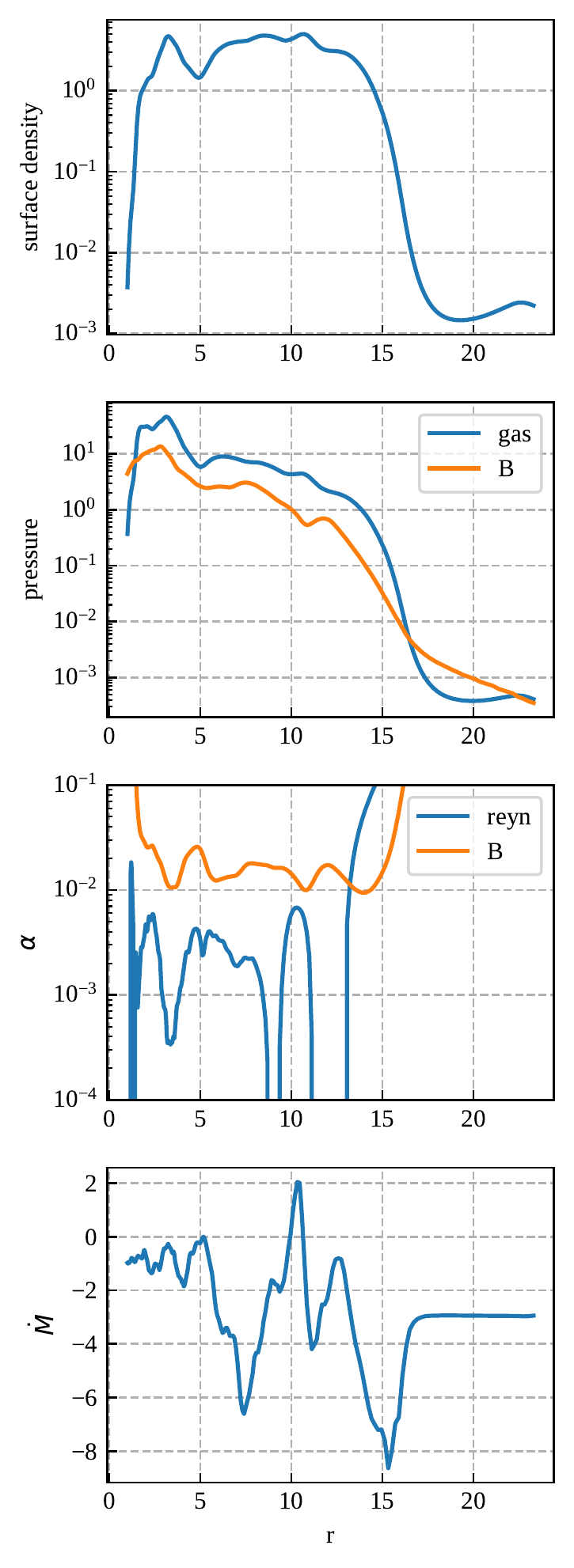}
    \caption{3D MHD midplane radial profiles at $t = 225$ binary orbits, time averaged over one binary orbit.  Negative $\dot M$ indicates accretion here.}
    \label{fig:mhd_3d_radial}
\end{figure}

Radial profiles of the 3D MHD simulation are plotted in Figure~\ref{fig:mhd_3d_radial}. The pressure is computed as the volume-weighted shell average.  The effective $\alpha$ is computed as
\begin{subequations}
    \begin{gather}
        \alpha = \frac{\langle T_{r\phi} \rangle}{\langle P \rangle}\label{eq:alpha_def}
    \end{gather}
where $\langle \cdot \rangle$ denotes shell averaging and $T_{r\phi}$ is the stress
    \begin{align}
        \text{Reynolds } T_{r\phi} &= \rho v_r (v_\phi - \frac{1}{2\pi}\int_0^{2\pi} d\phi\, v_\phi)\\
        \text{magnetic } T_{r\phi} &= -B_r B_\phi
    \end{align}
\end{subequations}
We note that the magnetic stress as measured by $\alpha$ is fairly constant at around $10^{-2}$ in the midplane for most radii of the disc. Magnetic $\alpha$ values of $\approx 10^{-3}$ to $10^{-2}$ were also seen in the simulations of \citet{patryk} in the midplane of their disks. Our $\dot M$ plot shows large variations in the outer radii of the disc, indicating the outer regions may not be in inflow equilibrium yet, even after 225 binary orbits. We can estimate the viscous time in the outer parts of the disc as
\begin{gather}
    t_\text{visc} = \frac{1}{\alpha \Omega(R)} \left( \frac{R}{H} \right)^2 \sim 1000 \text{ binary orbits}
\end{gather}
around $R = 10$ using $\alpha \sim 10^{-2}$ which is much longer than our simulation time with the stream. The large computational cost to evolve a disc for a viscous time presents a significant challenge for these MHD simulations, even in compact binary AM~CVn systems.

\begin{figure}
    \includegraphics[width=0.95\columnwidth]{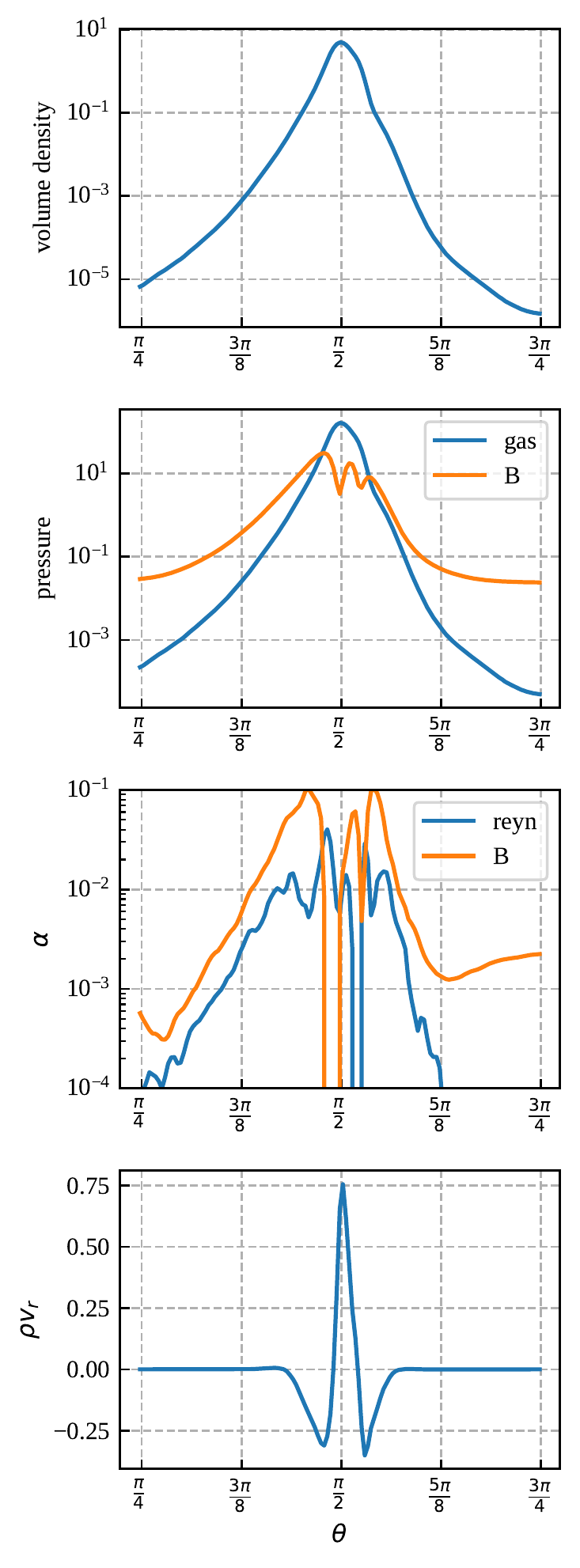}
    \caption{3D MHD azimuthally averaged vertical profiles at $r=6$ and at $t = 225$ binary orbits, time averaged over one binary orbit.}
    \label{fig:mhd_3d_angular}
\end{figure}

Vertical profiles as functions of polar angle $\theta$ near the binary circularization radius $r = 6$ are plotted in Figure~\ref{fig:mhd_3d_angular}. We defined $\alpha$ as before in Equation~(\ref{eq:alpha_def}). We see that the disc is gas pressure dominated in the midplane, but becomes magnetically dominated at altitude, consistent with many previous vertically stratified shearing box and global simulations of MRI turbulence \citep{sto96,mil00,haw01}.  The magnetic pressure profile rises as we approach the midplane from altitude but then flattens out in the midplane, suggestive of magnetic buoyancy. This behavior is also seen in the simulations of \citet{patryk}. The effective magnetic $\alpha$ shows 2 peaks at altitude away from the midplane, which coincides with the actual accretion occurring at these altitudes. This is reminiscent of the surface accretion that sometimes occurs in magnetically dominated high altitude layers in simulations with net poloidal field \citep{sto94,bec09,zhu18}, although our simulations here do not have net poloidal field.  The two-peaked profile of magnetic stresses is also seen at certain radii in \citet{patryk} in their Mach 5 simulation.

\section{Comparison of eccentricity evolution between 3D MHD and 2D simulations}\label{sec:eccent}
In this section we explore the evolution of eccentricity in our simulations in detail. As discussed in Section~\ref{sec:general_3d}, for the 3D MHD simulation, magnetic field loops were added into the disc at $t = 226$ binary orbits and the stream turned off in order to give it the best chance for producing eccentricity growth. As detailed in Section~\ref{sec:2d_setup}, the 2D alpha disc simulations are initialized using the data from the 3D MHD simulation at $t = 226$ binary orbits for a comparison and study of the eccentricity growth mechanisms. The three values of $\alpha$ used for comparison were $\alpha=0.01, 0.1, 0.2$. These 2D simulations are named alpha0.01, alpha0.1, alpha0.2 respectively.

We first present our methodology for analyzing eccentricity growth in simulations. Then we apply them to the 2D and 3D simulations.

\subsection{Diagnostics for eccentricity}\label{sec:eccent_method}
The Laplace-Runge-Lenz vector is a well-known conserved quantity in the
Kepler problem, where the acceleration is $\mathbf{a} = -GM_1
\mathbf{\hat{r}} / r^2$. Here we define a rescaled Laplace-Runge-Lenz vector,
the eccentricity vector, which has magnitude equal to the eccentricity and is
parallel to the semi-major axis, pointing towards periapsis.
\begin{equation}
\mathbf{e} = \frac{1}{GM_1} \mathbf{v} \times (\mathbf{r} \times \mathbf{v}) - \mathbf{\hat{r}} \label{eq:evec}
\end{equation}
Its time-derivative is given by
\begin{equation}
\frac{d \mathbf{e}}{dt} = \frac{1}{GM_1} (\mathbf{a} \times (\mathbf{r} \times \mathbf{v})
+ \mathbf{v} \times(\mathbf{r} \times \mathbf{a})) - \frac{d \mathbf{\hat{r}}}{dt} \label{eq:dedt}
\end{equation}
In an accretion disc in a binary system, additional forces such as pressure, tidal, and magnetic
would act on fluid elements so that $\mathbf{e}$ for fluid elements will not be
conserved in either magnitude or direction.

We construct the fluid analogue of Equation~(\ref{eq:dedt}). The
mass and momentum conservation equations solved by Athena++ can be equivalently written as
\begin{subequations}
    \begin{align}
        \partial_t \rho + \nabla \cdot (\rho \mathbf{v}) = 0 \label{eq:mass}\\
        \partial_t \mathbf{v} + \mathbf{v} \cdot \nabla \mathbf{v} = \mathbf{f}/\rho \label{eq:vel}
    \end{align}
\end{subequations}

We define the fluid eccentricity field $\mathbf{e}$ exactly the same as in
Equation~(\ref{eq:evec}) but with its symbols replaced by their respective fluid
fields. Noting that as field quantities $\partial_t \mathbf{\hat{r}} = 0$ and
$(\mathbf{v} \cdot \nabla \mathbf{r}) \times \mathbf{v} = 0$, we have using
Equation~(\ref{eq:vel})
\begin{gather*}
    \partial_t \mathbf{e} = \frac{1}{GM_1}(\mathbf{f}/\rho \times (\mathbf{r} \times \mathbf{v})
    + \mathbf{v} \times (\mathbf{r} \times \mathbf{f}/\rho))
    - \mathbf{v} \cdot \nabla \mathbf{e} - \mathbf{v} \cdot \nabla \mathbf{\hat{r}}
\end{gather*}
Multiplying both sides by $\rho$ and using Equation~(\ref{eq:mass}) we get
\begin{align}
    \partial_t (\rho \mathbf{e}) + \nabla \cdot (\mathbf{v} \rho \mathbf{e})
    &= \frac{1}{GM_1}(\mathbf{f} \times (\mathbf{r} \times \mathbf{v})
    + \mathbf{v} \times (\mathbf{r} \times \mathbf{f}))\nonumber\\
    &- \rho \mathbf{v} \cdot \nabla \mathbf{\hat{r}} \label{eq:fluid_evec}
\end{align}
which is the fluid version of Equation~(\ref{eq:dedt}).  The left-hand
side has the usual conservative form for per-volume fluid quantities, where
the second term represents advection by fluid motion. The right-hand side
consists of source terms.

We define the first term on the right-hand side as
\begin{gather}
    \mathbf{C}(\mathbf{f}) = \frac{1}{GM_1}(\mathbf{f} \times (\mathbf{r} \times \mathbf{v})
    + \mathbf{v} \times (\mathbf{r} \times \mathbf{f})) \label{eq:cvec_def}
\end{gather}
and remark that it is a linear function of $\mathbf{f}$. This linearity
allows us to split the source terms for eccentricity evolution into
individual parts each contributed by a force of interest (tidal, pressure,
etc). We also remark that as a consequence of the conservation of the
eccentricity vector in the Kepler problem
\begin{gather}
\mathbf{C}_\text{central gravity} (G M_1 \rho \mathbf{\hat{r}} / r^2)
- \rho \mathbf{v} \cdot \nabla \mathbf{\hat{r}} = 0 \label{eq:lenz_conserv}
\end{gather}
so neither of these terms needs to be computed in the analysis. Eccentricity
evolution is governed solely by those contributions to $\mathbf{C}$ due to the remaining
forces.

To measure the eccentricity evolution of the fluid in a certain volume in the simulation, we define the mass-weighted average eccentricity
\begin{gather}
    \langle \mathbf{e} \rangle_V = \frac{\int_V \rho\mathbf{e}\,dV}{\int_V \rho\,dV} = \frac{\int_V \rho\mathbf{e}\,dV}{M_\text{fluid}}\label{eq:sim_eccent_def}
\end{gather}
Its time evolution computed from Equation~(\ref{eq:fluid_evec}) is
\begin{align}
    \partial_t \langle \mathbf{e} \rangle_V &= \frac{1}{M_\text{fluid}} \left( \int_V \sum_\text{forces} \mathbf{C}(\mathbf{f})\,dV - \int_{\partial V} (\rho\mathbf{e}) (\mathbf{v}\cdot\mathbf{dA}) \right. \nonumber\\
    &\left. + \langle \mathbf{e} \rangle_V \int_{\partial V} \rho (\mathbf{v}\cdot\mathbf{dA}) \right)\label{eq:sim_eccent_evol}
\end{align}
where the summation is taken over all forces except the central gravity.

For both the 2D and 3D simulations, we take the argument of periapsis to be
\begin{gather}
    \varpi = \arctan\left(\frac{e_y}{e_x}\right)
\end{gather}
as computed in the non-rotating frame. This should be adequate for the 3D simulation as our disc is not tilted.

We define the growth and precession parts of $d\langle\mathbf{e}\rangle_V/dt$ as related to the components parallel and perpendicular to $\langle\mathbf{e}\rangle_V$ respectively
\begin{subequations}
    \begin{gather}
        \text{growth} = \left. \frac{d\lvert \mathbf{e} \rvert}{dt} = \left( \mathbf{e} \cdot \frac{d\mathbf{e}}{dt} \right) \middle/ \sqrt{\mathbf{e} \cdot \mathbf{e}} \right. \label{eq:eccent_growth}\\
        \text{prec} = \frac{d\varpi}{dt} = \left. \left( e_x \frac{d e_y}{dt} - e_y \frac{d e_x}{dt} \right) \middle/ (e_x^2 + e_y^2) \right. \label{eq:eccent_prec}
    \end{gather}
\end{subequations}
where we have written $\langle\mathbf{e}\rangle_V$ as $\mathbf{e}$ above for brevity. We remark that the right-hand side of these are linear functions of $d\langle\mathbf{e}\rangle_V/dt$ and therefore can be split into individual parts each contributed by a term of interest in Equation~(\ref{eq:sim_eccent_evol}).

\subsection{Diagnostics for excitation of eccentricity by spiral density waves}\label{sec:wave_method}
In perturbative treatments of eccentric discs, eccentricity is typically expressed as an $m=1$ perturbation of the form $e^{i \phi}$ in the nonrotating frame. For example, velocity perturbations for an eccentric disc are
\begin{subequations}
    \begin{align}
        \delta v_r &= -i r \Omega(r) e(r) e^{i \phi}\\
        \delta v_\phi &= \frac{1}{2} r \Omega(r) e(r) e^{i \phi}
    \end{align}\label{eq:perturb_eccent}
\end{subequations}
with $e(r)$ being the eccentricity magnitude and $\Omega(r)$ being the angular velocity for a circular disc.

In the mechanism of \citet{lubow_theory} for eccentricity growth, the gravity of the companion launches spiral density waves of the form $e^{i(m \pm 1)\phi - im\Omega_p t}$ in an eccentric disc. These waves further couple with tidal responses of the form $e^{im\phi - im\Omega_p t}$ to grow eccentricity in the form of Equation~(\ref{eq:perturb_eccent}).

In this paper, we have opted to define eccentricity through the Laplace-Runge-Lenz vector as it allows for a non-perturbative analysis of eccentricity evolution in simulation data through Equation~(\ref{eq:sim_eccent_evol}), applicable to the nonlinear regime. The downside is that we lose the perturbative picture in Equation~(\ref{eq:perturb_eccent}) with eccentricity as a function of radius and its relation to spiral waves through mode coupling.

Nevertheless, it is interesting to reconcile Lubow's wave mechanism with the present formalism to investigate the role of these spiral waves in simulation data. We work in two dimensions for simplicity. We can express the tidal contribution to eccentricity evolution with Equation~(\ref{eq:cvec_def}), rewritten as
\begin{align}
    \mathbf{C}_\text{tidal} &= \frac{r}{GM_1} (2 p_\phi a_\phi \mathbf{\hat{r}} - (p_r a_\phi + p_\phi a_r) \mathbf{\hat{\phi}}) \nonumber\\
    &= C_r \mathbf{\hat{r}} + C_\phi \mathbf{\hat{\phi}}
\end{align}
where $\mathbf{p} = \rho \mathbf{v}$ is momentum density, $\mathbf{a}$ is the tidal acceleration, and $C_r$ and $C_\phi$ are defined appropriately. For our companion moving in a circular orbit, $\mathbf{a}$ can be Fourier expanded in terms of the nonrotating frame tidal potential as
\begin{gather}
    \mathbf{a} = -\nabla\Phi = \sum_{m=-\infty}^\infty \left(-\frac{\partial \Phi_m}{\partial r} \mathbf{\hat{r}} - \frac{i m \Phi_m}{r} \mathbf{\hat{\phi}} \right) e^{i m \phi - i m \Omega_p t}
\end{gather}

To get the contribution to the global eccentricity growth, we should project $\mathbf{C}_\text{tidal}$ onto the normalized average eccentricity vector and normalize by the total fluid mass $M_\text{fluid}$ according to Equations~(\ref{eq:sim_eccent_evol}, \ref{eq:eccent_growth}). The result is
\begin{align}
    &\frac{\mathbf{C}_\text{tidal} \cdot \langle \mathbf{\hat{e}} \rangle_V}{M_\text{fluid}} = \frac{e^{i(\phi - \varpi)}}{M_\text{fluid}} \left( \frac{C_r + i C_\phi}{2} \right) + \text{c.c.} \nonumber\\
    &= \frac{1}{2 GM_1 M_\text{fluid}} \sum_{m=-\infty}^\infty e^{i (m + 1) \phi - i m \Omega_p t - i\varpi} \left(
    -m \Phi_m(2 i p_\phi + p_r) \right. \nonumber\\
    &\left. + i r p_\phi \frac{\partial \Phi_m}{\partial r}
    \right) + \text{c.c.}
\end{align}
with $\text{c.c.}$ denoting its complex conjugate. The above equation holds exactly in 2D as no approximations have been made so far. We see that the momentum density $\mathbf{p}$ is directly coupled to the tidal terms $\Phi$ to produce eccentricity growth, and therefore is the relevant quantity for analyzing spiral waves that grow eccentricity.

We can now introduce the approximations of $d\varpi/dt = 0$ and $dM_\text{fluid}/dt = 0$. These approximations hold if secular changes are slow compared to the orbital period. We choose our phase such that the companion is aligned with $\phi = 0$ at $t = 0$, so $\Phi_m$ is real. We can then integrate in azimuth and over one binary period to get the secular growth of eccentricity as a function of radius only. Then the Fourier components of $\mathbf{p}$ that survive are those with phase dependence $e^{-i(m+1)\phi + i m \Omega_p t}$ and its complex conjugate. For reference, the result is
\begin{align}
    \int_0^{\frac{2\pi}{\Omega_p}} &dt \int_0^{2\pi} d\phi\, \frac{\mathbf{C}_\text{tidal} \cdot \langle \mathbf{\hat{e}} \rangle_V}{M_\text{fluid}} \nonumber\\
    &= \frac{(2\pi)^2}{2 GM_1 M_\text{fluid} \Omega_p} e^{-i \varpi} \sum_{m=-\infty}^\infty
    \left(
    -m \Phi_m(2 i [p_\phi] + [p_r]) \right.\nonumber\\
    &\left. + i r [p_\phi] \frac{\partial \Phi_m}{\partial r}
    \right)
    + \text{c.c.}\label{eq:wave_analysis}
\end{align}
where $[\cdot]$ indicates only taking the $e^{-i(m+1)\phi + i m \Omega_p t}$ Fourier component
\begin{align}
    [f](r) = \int_0^{\frac{2\pi}{\Omega_p}} \frac{\Omega_p\,dt}{2\pi} \int_0^{2\pi} \frac{d\phi}{2\pi} f(t, r, \phi) e^{i(m+1)\phi - i m \Omega_p t}
\end{align}
Note that although we are analyzing eccentric discs, we have chosen to do our $\phi$ integration holding $r$ constant as opposed to integrating around an ellipse, since our simulation data naturally lies on a polar grid and leads to simpler analysis, without necessitating additional approximations.

Waves with $\lvert m+1 \rvert < \lvert m \rvert$ propagate inside inner eccentric Lindblad resonances whereas waves with $\lvert m+1 \rvert > \lvert m \rvert$ propagate outside outer eccentric Lindblad resonances. For our simulation of an AM CVn system, we expect that only the inner eccentric Lindblad resonances are relevant and are given approximately by the condition $\Omega = m \Omega_p / (m - 2)$ for $m > 0$ or equivalently $\Omega = m \Omega_p / (m + 2)$ for $m < 0$.

\subsection{Contributors to eccentricity evolution}
\begin{figure*}
    \begin{subfigure}{\columnwidth}
        \includegraphics[width=\columnwidth]{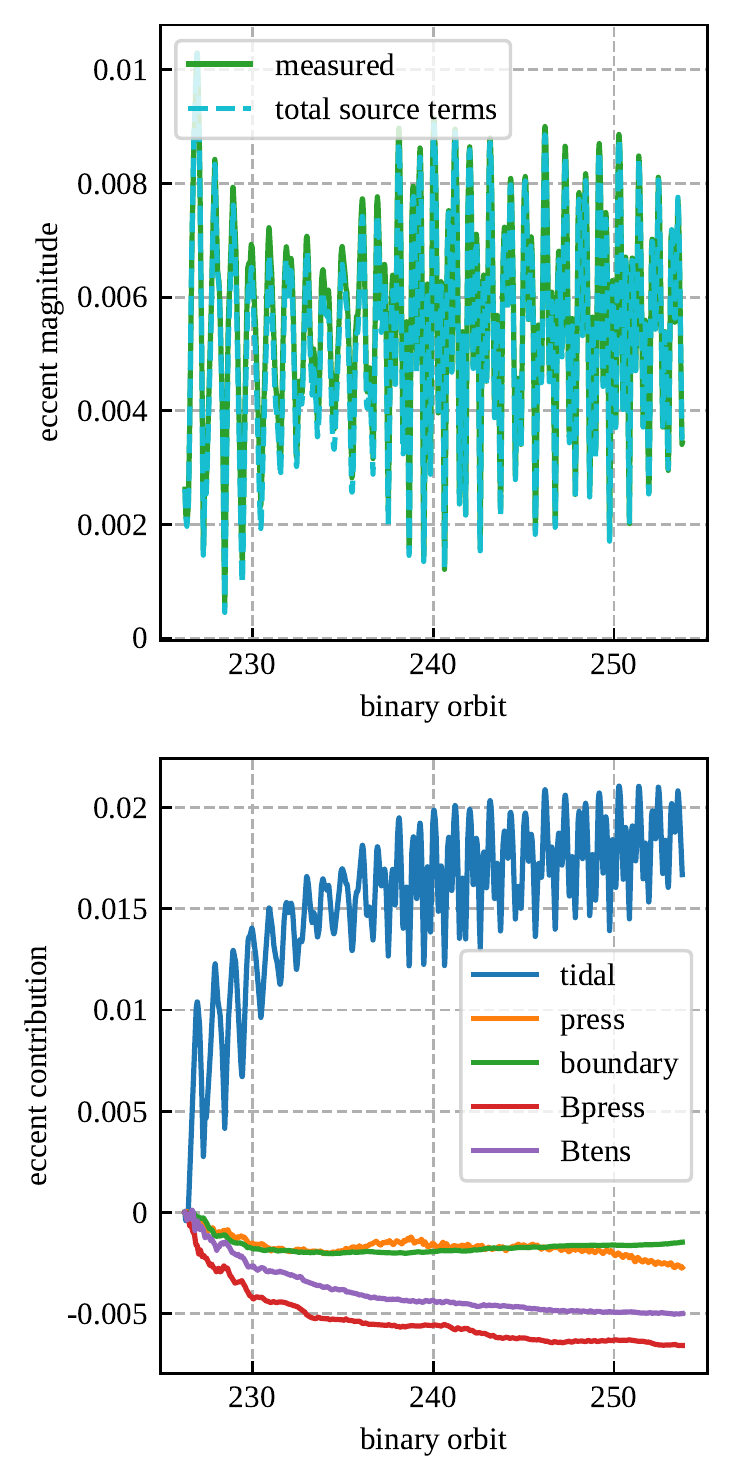}
    \end{subfigure}
    \begin{subfigure}{\columnwidth}
            \includegraphics[width=\columnwidth]{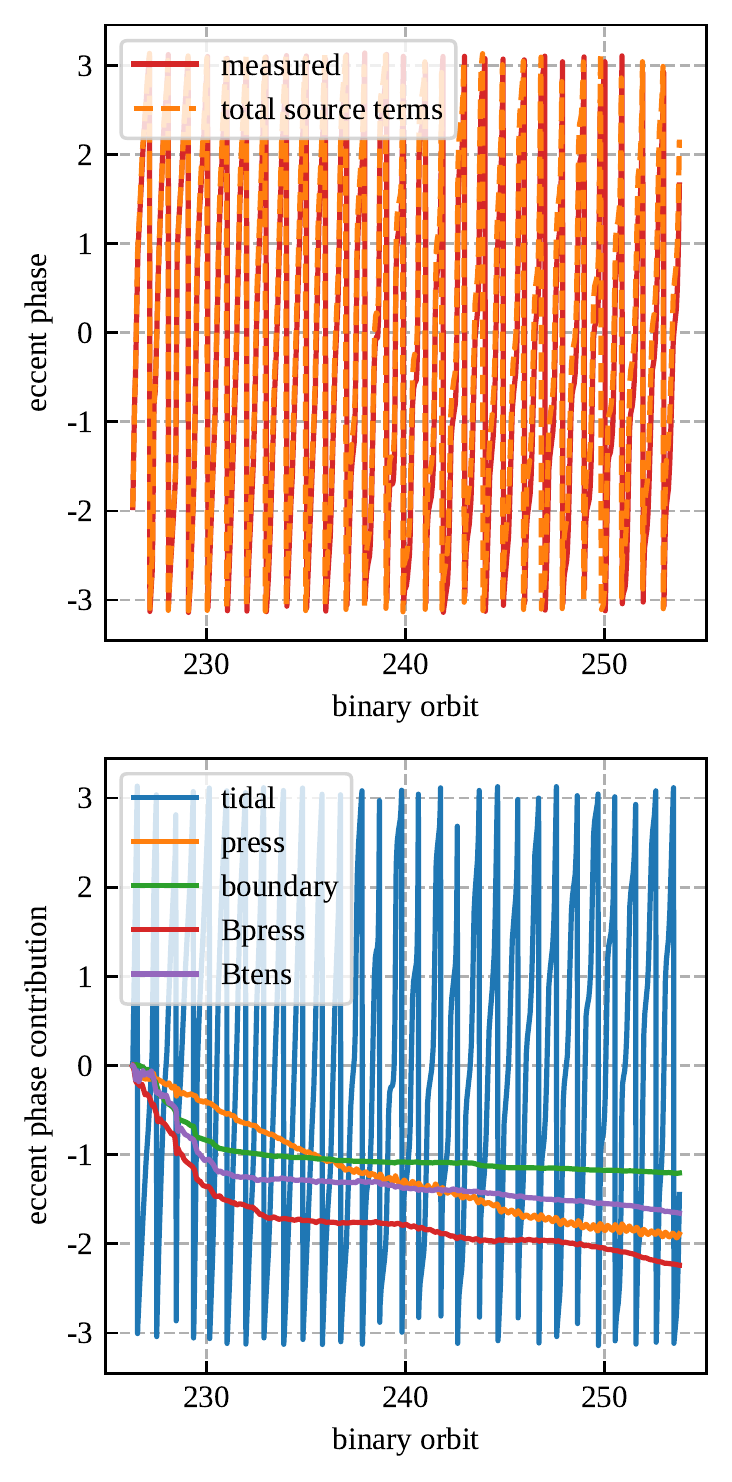}
    \end{subfigure}
    \caption{Eccentricity magnitude (left plots) and phase evolution (right plots) in 3D MHD simulation to compare the left-hand side (``measured'') and right-hand side (``total source terms'') of Equation (\ref{eq:sim_eccent_evol}), showing negligible eccentricity growth. Time integrated contribution of forces to eccentricity evolution are shown in bottom plots, with the sum of the curves in the bottom plots given in dashed curves in the top plots (``total source terms'' which include boundary effects), based on Equations (\ref{eq:sim_eccent_evol}, \ref{eq:eccent_growth}, \ref{eq:eccent_prec}). Matching of the dotted and solid lines in the top plots validates the numerical accuracy of the simulations and the methodology. The tidal force acts to increase eccentricity (blue curve, bottom left). Interestingly, the magnetic stresses act to decrease eccentricity (red and purple curves, bottom left). The eccentric phase $\varpi$ is measured with respect to a non-rotating frame centered on the primary. The rapid apsidal precession seen in the phase plots is a result of it tracking a non-wave pattern that follows the companion rather than the stationary ellipse pattern in the nonrotating frame needed for superhumps. The pressure force (orange curve, bottom right) contributes to retrograde apsidal precession. The boundary term (green curves, bottom plots) indicates the effect of fluid elements leaving the simulation domain as explained in the text.}
    \label{fig:mhd_3d_eccent_growth}
\end{figure*}

\begin{figure*}
    \begin{subfigure}{\columnwidth}
        \includegraphics[width=\columnwidth]{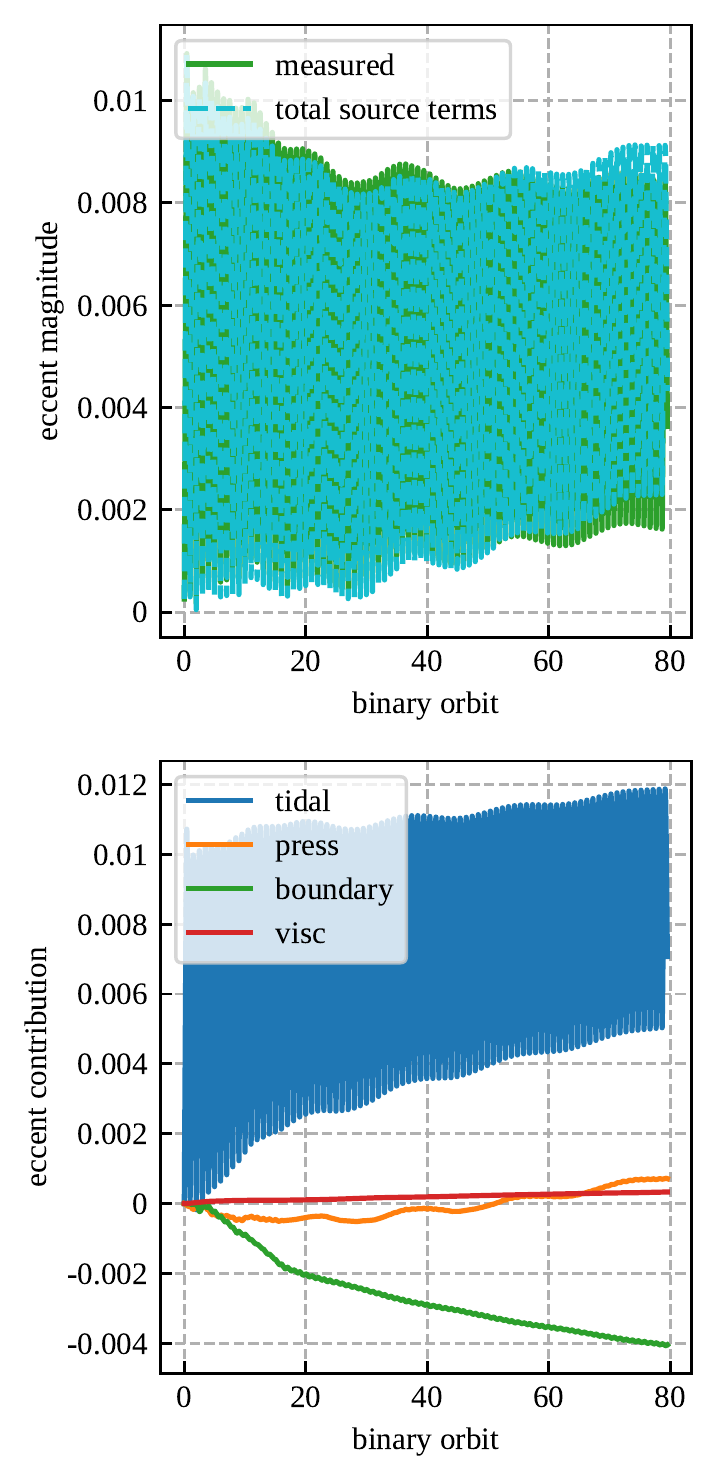}
    \end{subfigure}
    \begin{subfigure}{\columnwidth}
            \includegraphics[width=\columnwidth]{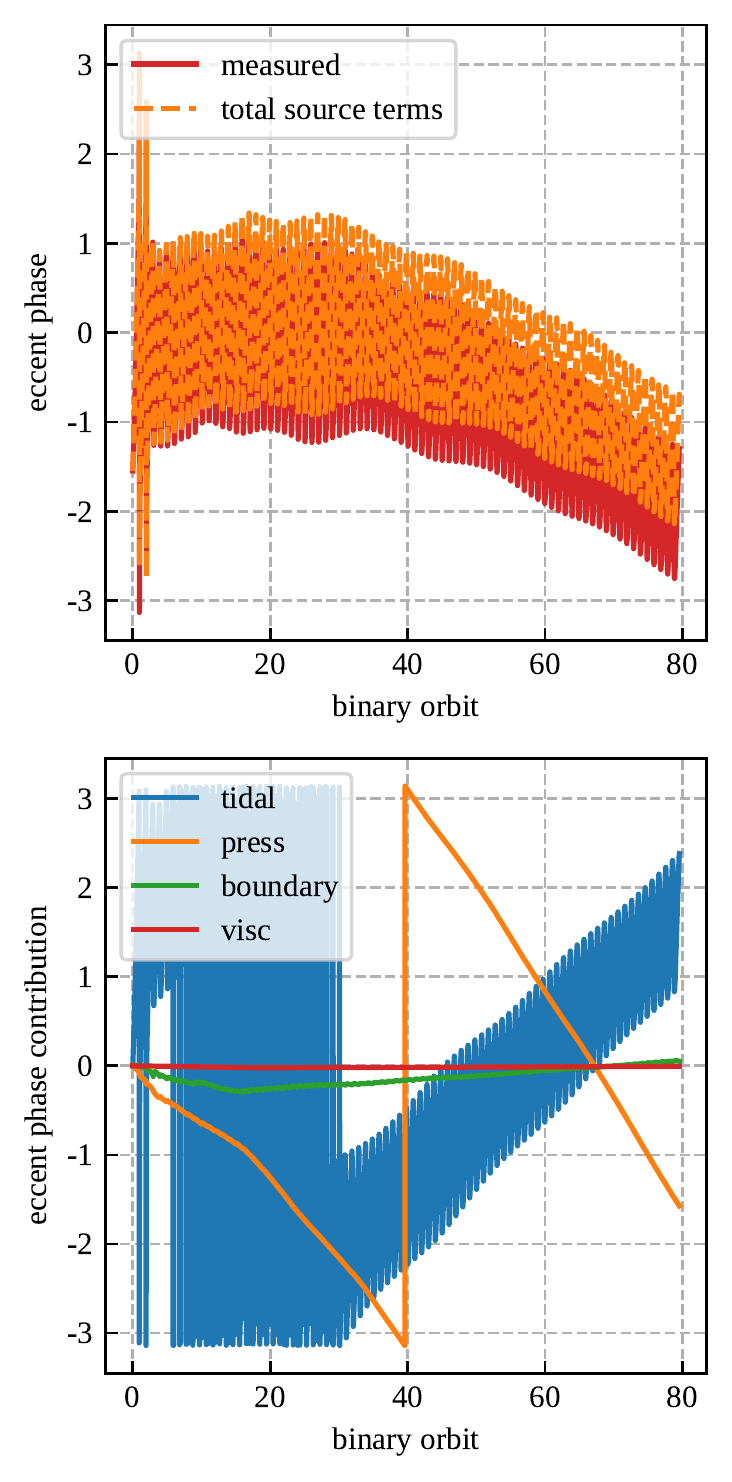}
    \end{subfigure}
    \caption{Eccentricity magnitude and phase evolution in 2D simulation with $\alpha=0.01$. See Figure~\ref{fig:mhd_3d_eccent_growth} for detailed description. The negligible eccentricity makes the phase $\varpi$ ill-defined for the right plots.}
    \label{fig:alpha0.01_eccent_growth}
\end{figure*}

\begin{figure*}
    \begin{subfigure}{\columnwidth}
        \includegraphics[width=\columnwidth]{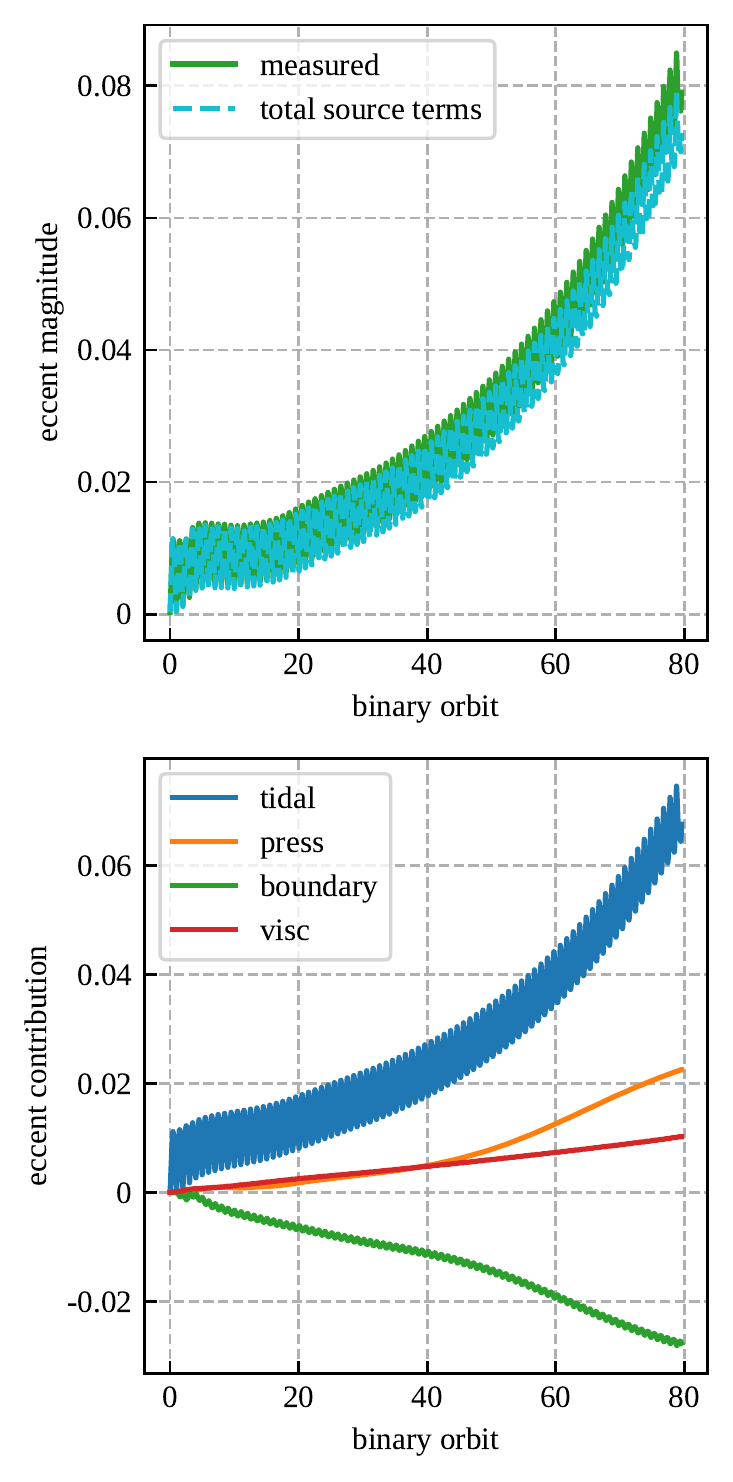}
    \end{subfigure}
    \begin{subfigure}{\columnwidth}
            \includegraphics[width=\columnwidth]{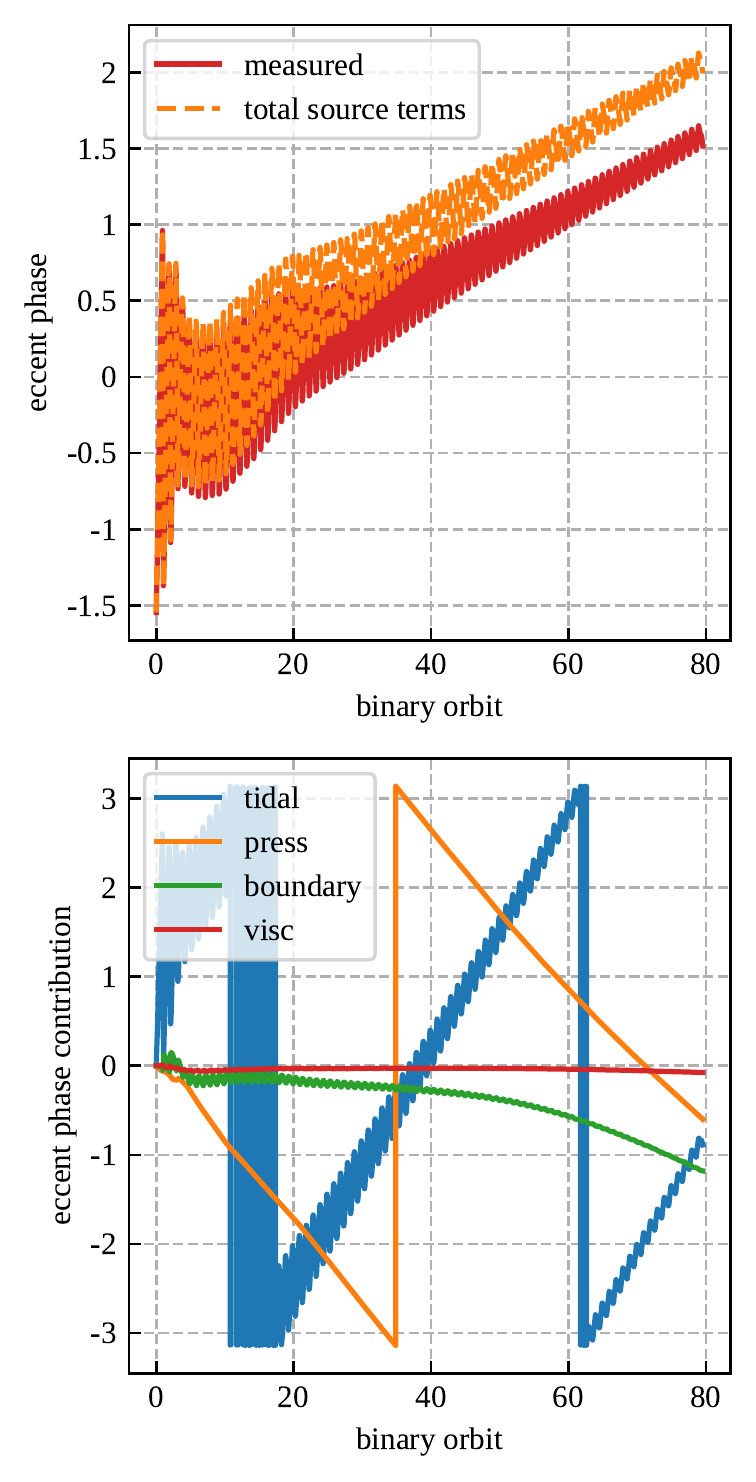}
    \end{subfigure}
    \caption{Eccentricity magnitude and phase evolution in 2D simulation with $\alpha=0.1$. See Figure~\ref{fig:mhd_3d_eccent_growth} for detailed description. Slower precession is seen compared to the $\alpha=0.2$ simulation.}
    \label{fig:alpha0.1_eccent_growth}
\end{figure*}

\begin{figure*}
    \begin{subfigure}{\columnwidth}
        \includegraphics[width=\columnwidth]{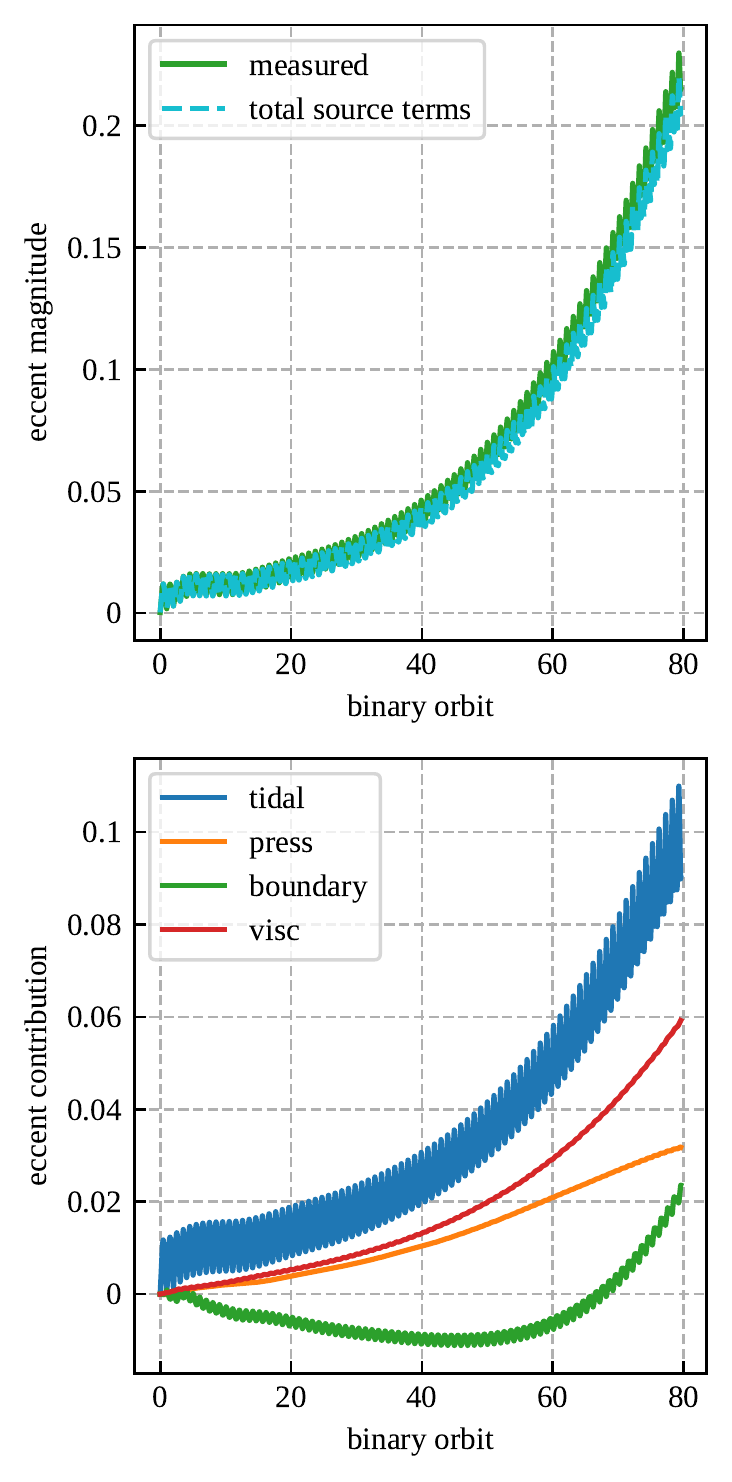}
    \end{subfigure}
    \begin{subfigure}{\columnwidth}
            \includegraphics[width=\columnwidth]{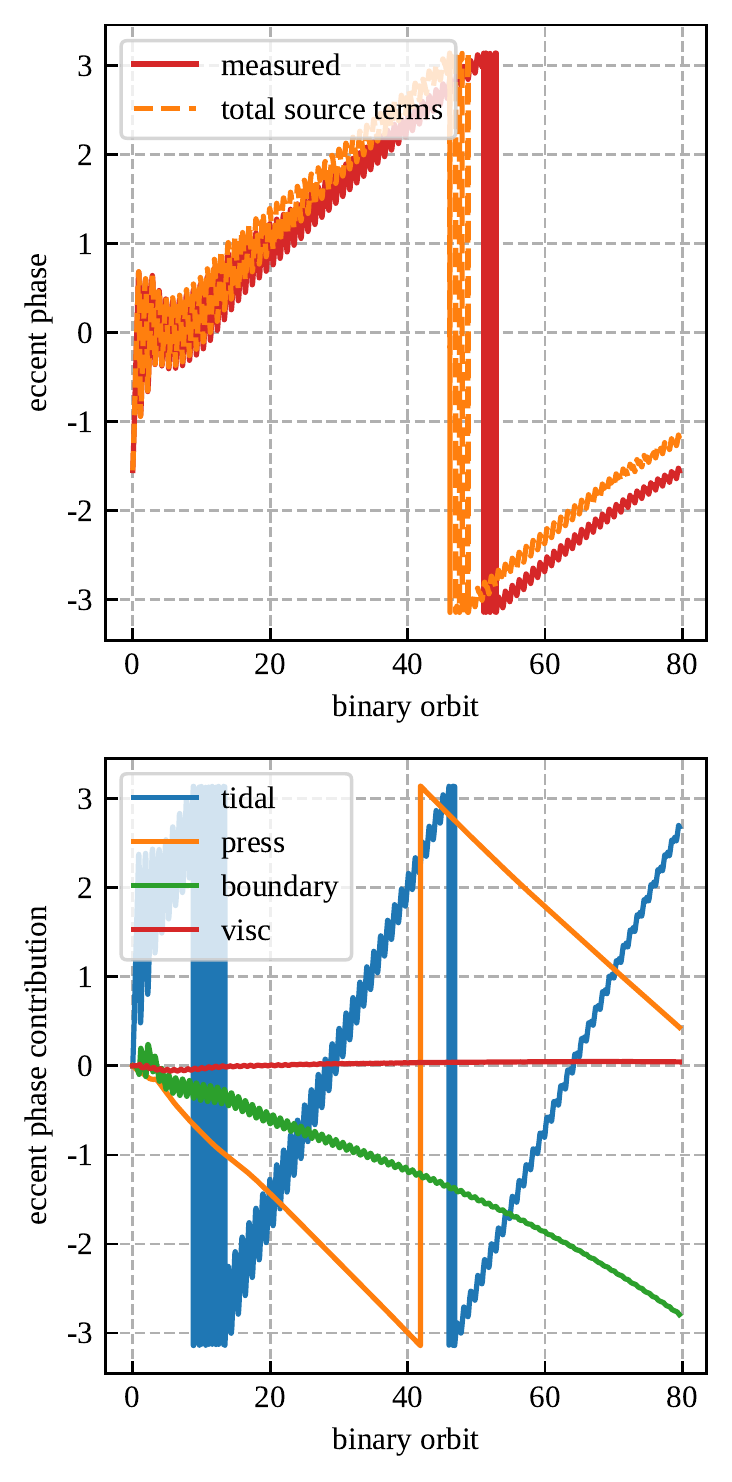}
    \end{subfigure}
    \caption{Eccentricity magnitude and phase evolution in 2D simulation with $\alpha=0.2$. See Figure~\ref{fig:mhd_3d_eccent_growth} for detailed description. More rapid precession is seen compared to the $\alpha=0.1$ simulation, but it is not driven by the viscous force (red curve, bottom right).}
    \label{fig:alpha0.2_eccent_growth}
\end{figure*}

Figure~\ref{fig:mhd_3d_eccent_growth} shows the evolution of eccentricity in the 3D MHD simulation after $t=226$ binary orbits. Figures \ref{fig:alpha0.01_eccent_growth}, \ref{fig:alpha0.1_eccent_growth}, and \ref{fig:alpha0.2_eccent_growth} show the evolution of eccentricity in three 2D simulations with alpha values of $\alpha = 0.01, 0.1, 0.2$. We use Equations (\ref{eq:sim_eccent_evol}, \ref{eq:eccent_growth}, \ref{eq:eccent_prec}) integrated in time to understand the role of the various forces on eccentricity evolution. We take our integration volume $V$ to be the entire simulation domain in both the 3D and 2D simulations. On the right-hand side of Equation~(\ref{eq:sim_eccent_evol}), we split the $\mathbf{C}(\mathbf{f})$ source term defined in Equation~(\ref{eq:cvec_def}) into contributions from the forces affecting the fluid in each simulation. The last two terms on the right-hand side, involving fluid entering or leaving the simulation boundaries, are combined into a single term called ``boundary'' (green curves in bottom plots) to account for flux terms through the inner and outer boundary.
In all simulations, there is good agreement between the left and right hand sides of Equation~(\ref{eq:sim_eccent_evol}), indicated by matching of the solid and dotted curves in the top plots, validating both the numerical accuracy and our methodology for quantifying eccentricity sources.

We do not observe significant eccentricity growth in the 3D MHD simulation. The tidal force initially acts to increase eccentricity (blue curve, bottom left) but its effect is canceled by the remaining sources. The effect of magnetic fields is to decrease eccentricity. This is noteworthy as the turbulent magnetic stresses are typically thought of as the physical mechanism behind the $\alpha$ viscosity, yet here the magnetic stresses act with a different sign on eccentricity growth compared to the $\alpha$ viscosity in the 2D simulations.

\begin{figure}
    \includegraphics[width=\columnwidth]{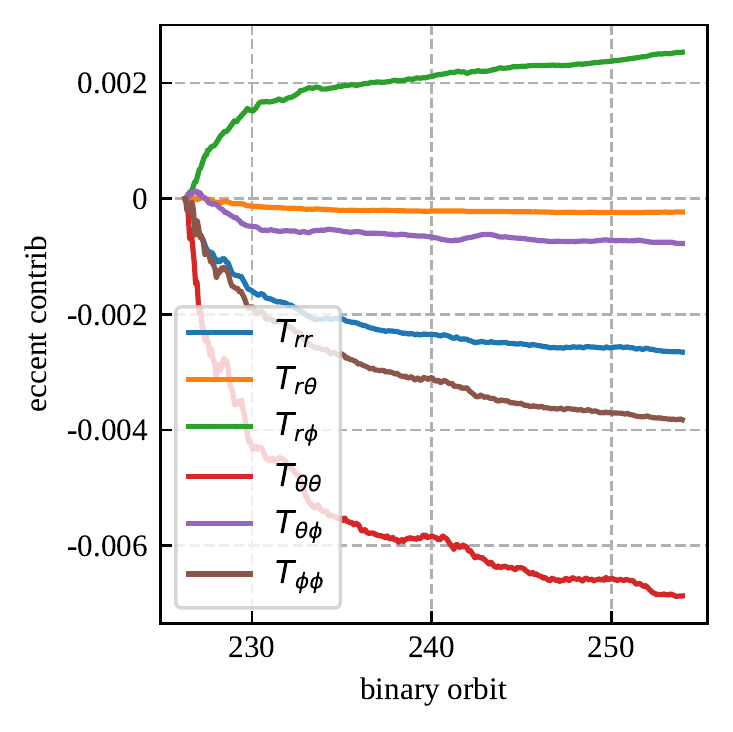}
    \caption{The time integrated contribution to eccentricity from the individual pieces of the Maxwell stress tensor.}
    \label{fig:mhd_3d_transport}
\end{figure}

In Figure \ref{fig:mhd_3d_transport}, we compute the time-integrated contribution to eccentricity growth of the individual pieces of the Maxwell stress tensor. Specifically, we zero out all components of the Maxwell stress tensor except the indicated component and its symmetric counterpart, and we compute the force with $\nabla \cdot T$, then use that force in Equation (\ref{eq:cvec_def}) to compute its effect on eccentricity. In particular, we find that the $B_r B_\phi$ acts to increase eccentricity whereas the $B_r B_\theta$ piece damps eccentricity. The $B_r B_\phi$ piece is responsible for the outward radial transport of angular momentum whereas the $B_\theta B_\phi$ piece is responsible for the vertical transport of angular momentum. In the 2D disc simulations, the $\alpha$ viscosity is responsible for the outward radial transport of angular momentum, and it is noteworthy that both the $B_r B_\phi$ and $\alpha$ viscosity act to increase eccentricity.


Of the three 2D simulations, significant eccentricity growth (top green curve) is only observed in the $\alpha=0.1$ and $0.2$ cases. The larger viscosity simulations have more rapid eccentricity growth, replicating the result of \citet{kley}. The eccentricity growth is predominantly driven by tidal forces (bottom blue curve) as expected. \citet{lyubarskij_eccent} and \citet{ogilvie_nonlinear_eccent} have previously noted that circular $\alpha$-discs are unstable to eccentric modes. Here, we find that the viscous force (bottom red curve) also directly contributes to eccentricity growth, especially in the $\alpha=0.2$ case where it is half as large as the tidal contribution, but we do not know whether this is related to the mechanism in the analytic theories. While eccentricity is still small, the simulation boundaries (bottom green curve) contribute negatively to eccentricity growth, indicating fluid elements on eccentric orbits leaving the domain and thus removing eccentricity.

 Though both the 3D MHD and 2D hydro simulations are done in a rotating frame, the argument of periapsis $\varpi$, also called the eccentric phase here, is measured with respect to a non-rotating frame centered on the primary.

In the $\alpha=0.1$ and $0.2$ simulations, we can see that the only two significant contributors to the precession of the eccentric disc are the tidal and pressure forces. The tidal force contributes to prograde precession of the disc, while the pressure force usually contributes to retrograde precession of the disc. The effects of varying the disc temperature and hence the pressure force and disc scale height have been previously explored in other works, and it is known \citep{lubow92, goodchild_ogilvie,kley} that typical 2D discs precess in a prograde fashion for smaller disc scale heights and retrograde for large scale heights where pressure forces become significant, in agreement with our findings. Observationally, positive superhumps suggest overall prograde precession of discs, suggesting that discs in nature typically are not in the regime in which pressure effects on precession dominate over tidal effects.

Also noteworthy is that the $\alpha=0.2$ simulation experiences more rapid apsidal precession as compared to $\alpha=0.1$. The effect of larger viscosity resulting in more rapid apsidal precession has been previously noted in earlier SPH simulations \citep{mur98}. In our analysis we find that the direct effect of the viscous force on precession is negligible in both the $\alpha=0.1, 0.2$ simulations (red curves, bottom right plots). Since only the tidal force acts more strongly to precess the disc prograde in the $\alpha=0.2$ simulation, and the tidal acceleration is the same in both simulations, the cause must be the mass distribution of the disc. As further explored in Section~\ref{sec:density_distrib}, $\alpha=0.2$ has more mass in the outer regions of the disc as compared to the $\alpha=0.1$ simulation. This indicates that the precession rate and hence the superhump frequency should also depend on the mass distribution of the disc which is affected by the angular momentum transport mechanism.


Since we do not have a significant eccentricity in the 3D MHD simulation, the eccentric phase $\varpi$ is not well-defined. The phase plot for $\varpi$ is instead tracking a small non-wave $e^{i\phi - i\Omega_p t}$ pattern that follows the companion.  


\subsection{Wave analysis of the mode-coupling mechanism}
As indicated in the previous section, the tidal force is the dominant contributor to eccentricity growth in the simulations. In this section, we investigate the tidal effect on the disc in greater detail through Lubow's spiral wave mechanism by directly computing each wave's contribution to eccentricity growth from simulation data. In Lubow's theory, the tidal force acting on a disc with an initially small eccentricity would launch spiral waves in the disc, primarily driven at eccentric Lindblad resonances. These spiral waves are then further acted on by tides to generate additional eccentricity in the disc \citep{lubow_theory}.

In Section~\ref{sec:wave_method}, we discussed a method for quantifying the role of these waves in our simulations. We first measured the Fourier components of the momentum densities $p_r$ and $p_\phi$ from the simulation and then we coupled them to the tidal field with Equation~(\ref{eq:wave_analysis}) to compute the contribution of each relevant spiral wave to the overall eccentricity growth. This is done for each radius separately to look for the role of any resonances and to investigate the spiral nature of these waves.

An important assumption made for the validity of Equation~(\ref{eq:wave_analysis}) is that the argument of periapsis $\varpi$ varies little during a single binary orbit. Physically, we are only interested in eccentricity as defined by an elliptical disc stationary in a nonrotating frame, and want to ignore any $e^{i\phi - i\Omega_b t}$ tidal distortions that would follow and rotate along with the companion. The assumption of a stationary argument of periapsis $\varpi$ is found to not be valid when the eccentricity is small (see for example the early times in Figure~\ref{fig:alpha0.1_eccent_growth}), since a small eccentricity vector can easily precess wildly in angle from comparably small perturbations. Eccentricity is always small for the 3D MHD and the $\alpha=0.01$ simulations, and is small in the early stages of the $\alpha=0.1$ and $\alpha=0.2$ simulations. Therefore, the wave analysis is only most useful in the latter stages of the $\alpha=0.1$ and $\alpha=0.2$ simulations where there is a well-defined eccentricity with slowly precessing argument of periapsis. Since the tidal field is the same among these simulations, we choose the $\alpha=0.1$ simulation for our main analysis of the wave coupling mechanism.

\begin{figure}
    \includegraphics[width=\columnwidth]{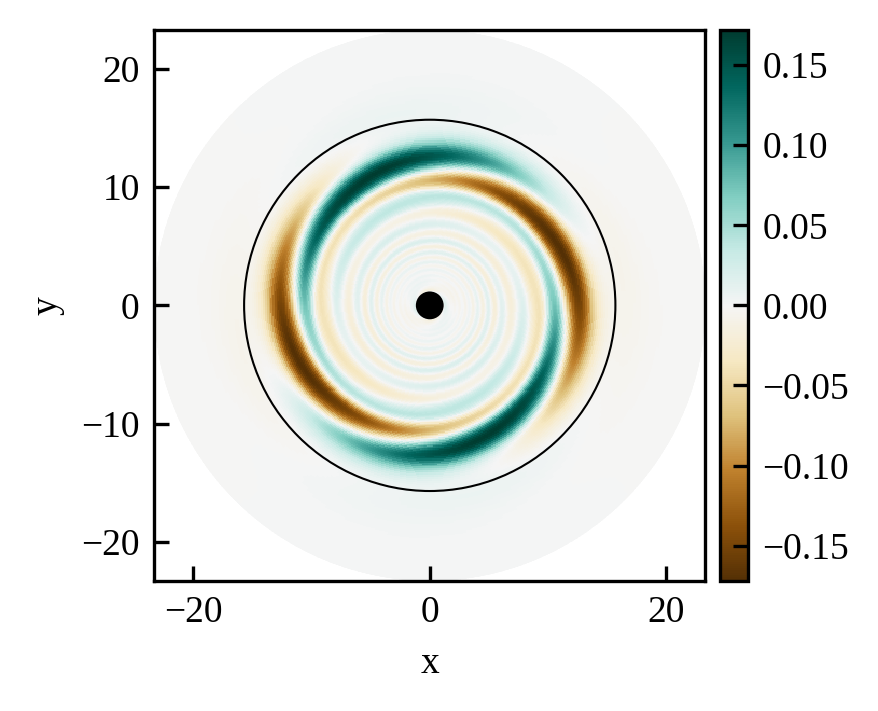}
    \caption{Real part of the spiral density wave in the radial momentum $p_r$ of the form $e^{i(2\phi - 3\Omega_p t)}$ measured from the 2D $\alpha = 0.1$ simulation at $t = 79$ binary orbits. Its eccentric Lindblad resonance is the binary's 3:1 resonance located at $r = 15.2$, indicated by black circle. Though the wave amplitude becomes small at larger radii, its contribution to the globally averaged eccentricity growth is not negligible there because the tidal force is also larger (see Figure~\ref{fig:2d_radial_wave} orange curve)}
    \label{fig:2d_lubow_wave}
\end{figure}

We use the notation $(n, l)$ to indicate waves of the form $e^{i n \phi - i l \Omega_p t}$ in the non-rotating frame centered on the primary. Figure~\ref{fig:2d_lubow_wave} shows a plot of the real part of the $(2, 3)$ wave for $p_r$ as measured from the simulation. The $(2, 3)$ wave is significant because it is thought to be the primary wave responsible for eccentricity growth when it is excited by the 3:1 eccentric Lindblad resonance at $r = 15.2$.

\begin{figure}
    \includegraphics[width=\columnwidth]{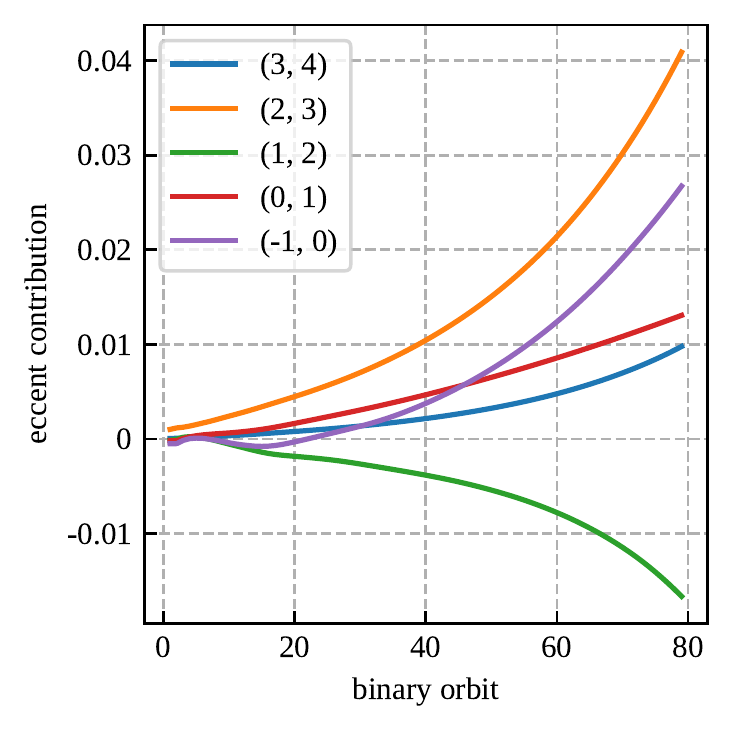}
    \caption{Time-integrated contribution to eccentricity growth due to spiral waves coupled with the tidal force, via Lubow's mechanism. $(n, l)$ indicates spiral waves of the form $e^{i n \phi - i l \Omega_p t}$. Largest contribution to eccentricity growth comes from the (2, 3) wave (orange curve) excited by the 3:1 resonance, but there is significant contribution from other waves as well.}
    \label{fig:2d_time_wave}
\end{figure}

In Figure~\ref{fig:2d_time_wave} we compute the time and spatial integral of the right-hand side of Equation~(\ref{eq:wave_analysis}) over many binary orbits to determine the total long-term secular effect of each wave on global eccentricity growth. We confirm that the $(2, 3)$ wave (orange curve) is the dominant contributor to eccentricity growth, but that other waves also contribute significantly. It is interesting that the $(-1, 0)$ ``wave'' has the second largest contribution to eccentricity growth over time.

\begin{figure}
    \includegraphics[width=\columnwidth]{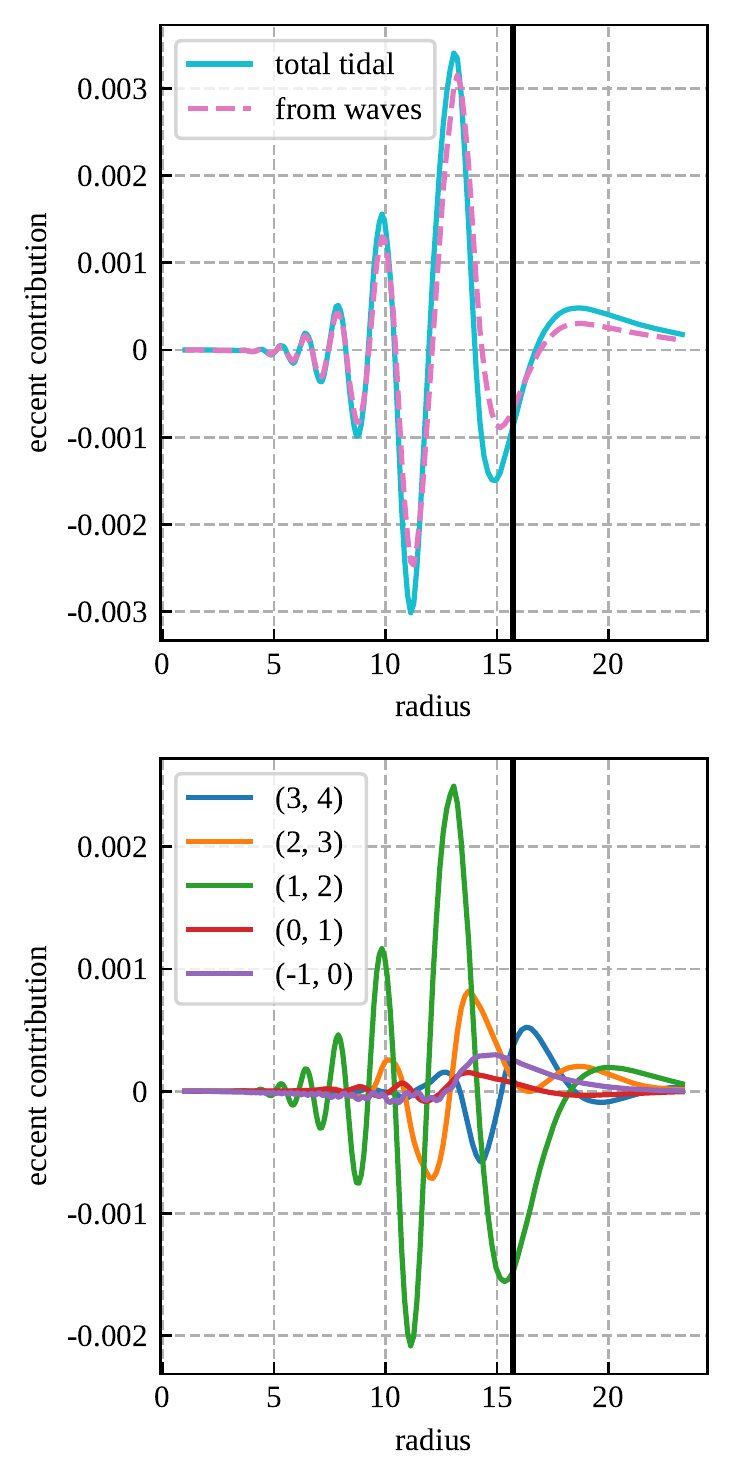}
    \caption{Radial dependence of the contribution to eccentricity growth due to spiral waves coupled with the tidal force, via Lubow's mechanism, at $t = 79$ binary orbits. In the top plot, ``total tidal'' (solid line) refers to the total effect of the tidal force on eccentricity growth (LHS of Equation~(\ref{eq:wave_analysis})), and the ``from waves'' (pink dotted line) is the sum of the curves from the bottom plot only, which only includes the Fourier modes shown. Bottom plot shows the contributions of each mode to eccentricity growth (RHS of Equation~(\ref{eq:wave_analysis})). Black vertical line indicates the binary's 3:1 resonance. $(n, l)$ indicates spiral waves of the form $e^{i n \phi - i l \Omega_p t}$. We have multiplied by an additional factor of $r$ compared to Equation~(\ref{eq:wave_analysis}) so that the radial integral of these plots with measure $dr$ gives the total eccentricity growth contribution.}
    \label{fig:2d_radial_wave}
\end{figure}

Figure~\ref{fig:2d_radial_wave} shows the right-hand side of Equation~(\ref{eq:wave_analysis}) plotted as a function of radius. In the top plot we first confirm that the left-hand side of Equation~(\ref{eq:wave_analysis}) (solid line), which consists of time-integrating the instantaneous tidal effect on eccentricity growth, matches the right-hand side (dotted line), which uses the decomposition into waves and makes the approximations $d\varpi/dt = 0$ and $dM_\text{fluid}/dt = 0$. We also only sum over the Fourier modes shown in the bottom plot to arrive at the dotted line, which we found to be the only non-negligible modes. The good agreement between the solid and dotted line in the top plot validates the numerical accuracy of our simulation and the approximations made in Equation~(\ref{eq:wave_analysis}).

Several features of these waves are noteworthy. First, the radial oscillations in the contribution to global eccentricity growth indicates the spiral nature of these waves, which is also readily seen in Figure~\ref{fig:2d_lubow_wave}. Spiral waves wind in azimuth as we move radially, so for roughly half the radii they will have the wrong phase for eccentricity growth and will instead damp the globally averaged eccentricity. However, in theory, they should have a consistent phase near their associated Lindblad resonance since the radial WKB wavenumber approaches 0, and they also have a consistent phase in their evanescent region on one side of resonance. So we expected the most important wave, the $(2, 3)$ wave (orange curve), to become evanescent beyond its Lindblad resonance, the 3:1 resonance, at $r = 15.2$. However, the transition to evanescence is not as expected since oscillations are still seen beyond the resonance, albeit not through zero. The cause and implication of this behavior is not clear to us, though it may be related to viscous or nonlinear effects.

Next, though the amplitude of the $(2, 3)$ spiral wave as seen in Figure~\ref{fig:2d_lubow_wave} appears to be negligible beyond the 3:1 resonance, its contribution to eccentricity growth is not negligible at larger radii as seen in the orange curve of Figure~\ref{fig:2d_radial_wave}. This is because the contribution to eccentricity growth has an additional coupling with the tidal field $\Phi$ which is also larger for larger radii.

The largest amplitude in radial oscillation of the contribution to eccentricity growth in Figure~\ref{fig:2d_radial_wave} is the $(1, 2)$ wave (green curve). However, its long-term contribution to eccentricity growth when integrated over all radii is negative as seen in Figure~\ref{fig:2d_time_wave}, and is still smaller in absolute value compared to the $(2, 3)$ wave. This is because in the oscillatory region we get cancellation between neighboring peaks with opposite sign.

\begin{figure}
    \includegraphics[width=\columnwidth]{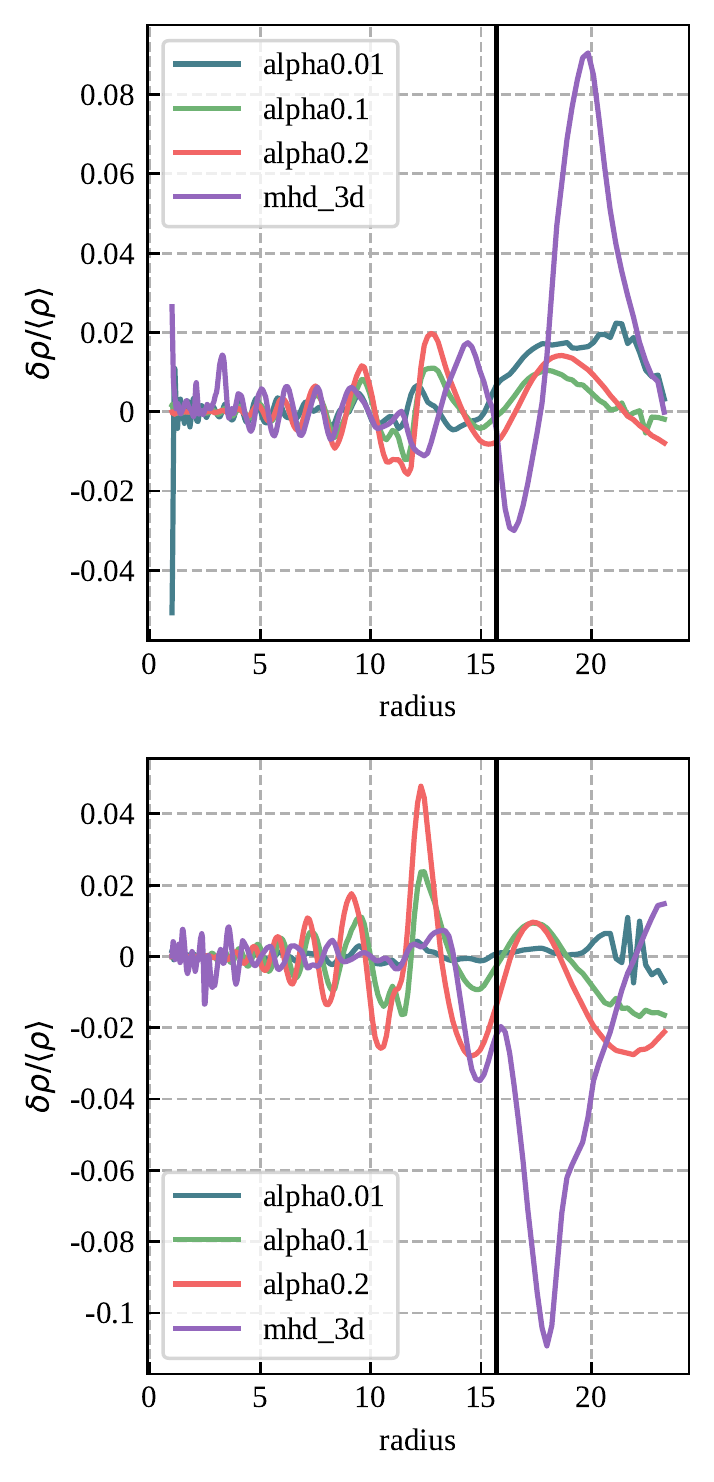}
    \caption{Comparison of the $e^{i(2\phi - 3\Omega_p t)}$ normalized density wave across simulations at different times: top plot is at 5 binary orbits, bottom plot is at 20 binary orbits. $\delta\rho$ is the real part of the wave given by Equation~(\ref{eq:dens_wave}), and $\langle \rho \rangle$ is the azimuthally averaged density. The amplitude increases over time for the higher $\alpha$ 2D simulations, but has decreased in amplitude for the 3D MHD simulation in $r < 12$. Vertical black line indicates the location of its 3:1 eccentric Lindblad resonance. The top and bottom curves for the 3D MHD simulation are taken at 5 and 20 binary orbits, respectively, after $t = 226$ binary orbits to allow comparisons with the 2D simulations.}
    \label{fig:dens_wave_compare}
\end{figure}

Since the analysis of the waves in the 2D $\alpha=0.1$ simulation show that the $(2, 3)$ wave driven by the 3:1 resonance is the most important for eccentricity growth, we wish to look for this wave in the 3D simulation. As explained earlier in this section, we cannot compute the wave contribution to eccentricity growth in the 3D MHD simulation since it does not have a well-defined eccentricity vector to project onto. Instead, we simply compute the relative amplitudes of the $(2, 3)$ density wave in the 3D and 2D simulations for comparison. Density was chosen over the momenta $p_r$, $p_\phi$ used above only for simplicity. We compute the wave amplitude in the 3D simulation by first vertically integrating in $\theta$ then picking the real part of the $(2, 3)$ Fourier component, resulting in
\begin{gather}
    \delta \rho = \text{Re} \left[ \int_{t_0}^{t_1} \frac{\Omega_p\,dt}{2\pi} \int_0^{2\pi} \frac{d\phi}{2\pi} \int_{0}^{\pi} d\theta\, \rho e^{-i 2 \phi + i 3 \Omega_p t} r \sin \theta \right]\label{eq:dens_wave}
\end{gather}
with the time integration being over one orbit. We likewise compute the wave amplitude in the 2D simulation without the $\theta$ vertical integration. $\delta\rho$ is then normalized by the azimuthally averaged density to enable comparisons between simulations and plotted in Figure~\ref{fig:dens_wave_compare}.


We see that in the inner regions of the disc where $r < 12$, the amplitude of the $(2, 3)$ density wave in the 3D simulation (purple curve) decreases over time. This coincides with the shrinking of the disc as described in Section~\ref{sec:density_distrib}. As less material reaches the 3:1 resonance over time, the most important wave contributing to eccentricity growth diminishes in amplitude. We also see evidence for this in Figure \ref{fig:mhd_3d_eccent_growth}, where the tidal force initially contributes to eccentricity growth before ceasing to do so. On the other hand, the 2D simulations show an increase in the amplitude of the $(2, 3)$ density wave over time. This is caused both by the spreading of the disc over time allowing for more mass at the 3:1 resonance, and also the increased eccentricity at later time coupling more strongly to the tides to produce the $(2, 3)$ wave.

\subsection{Surface density evolution}\label{sec:density_distrib}
\begin{figure}
    \includegraphics[width=\columnwidth]{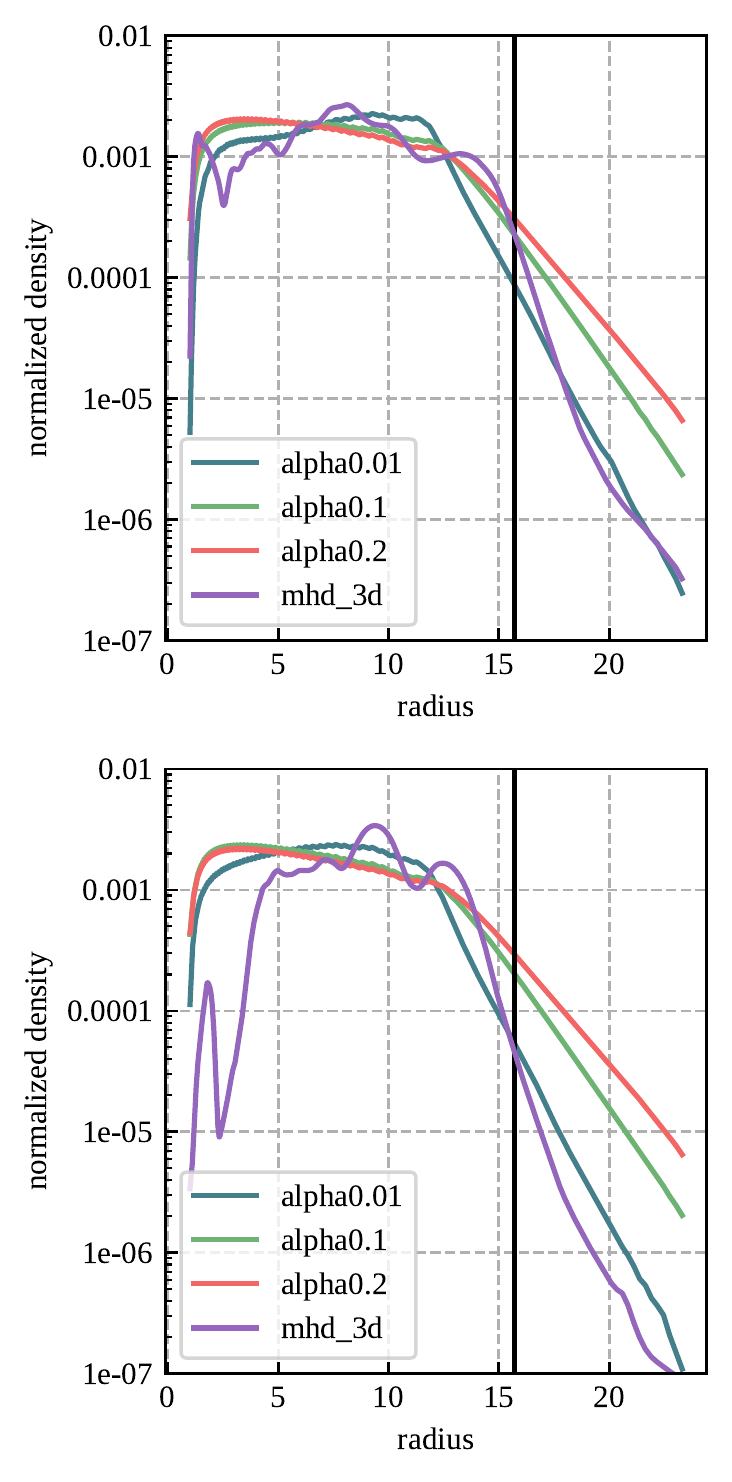}
    \caption{Azimuthally averaged surface density radial profiles, normalized by total mass: top plot is at 5 binary orbits, bottom plot is at 20 binary orbits. The surface density of the 3D simulation at the 3:1 resonance is comparable to the larger $\alpha$ 2D simulations shortly after the magnetic field loops are added, but then decreases over time, coinciding with the diminished tidally driven eccentricity growth (see Figure~\ref{fig:mhd_3d_eccent_growth}). Tidal truncation by the companion gravity counters the turbulent/viscous spreading of the disc. Vertical black line indicates the location of the nominal 3:1 resonance. The top and bottom curves for the 3D MHD simulation are taken at 5 and 20 binary orbits, respectively, after $t = 226$ binary orbits to allow comparisons with the 2D simulations.}
    \label{fig:surf_dens_compare}
\end{figure}

Figure~\ref{fig:surf_dens_compare} shows the normalized and azimuthally averaged surface density of the simulations at two different times. The surface densities were normalized by the instantaneous total mass in the simulations to enable relative comparisons between them, since some simulations lose mass more rapidly. We see that only after around 5 binary orbits, the surface densities of all three 2D simulations look similar to the surface density at the much later time of around 30 binary orbits. The readjustment of surface density correlates with the eccentricity growth plots in Figures \ref{fig:alpha0.1_eccent_growth}, \ref{fig:alpha0.2_eccent_growth}, where we see an initial transient behavior within the first 5 binary orbits before entering the exponential growth part. There is significantly more mass in the outer regions of the disc in the higher alpha simulations, and eccentricity growth only occurs in these simulations. This agrees with the result of \citet{kley} that the eccentricity growth can only take place after significant mass is present in the outer regions of the disc.

The 3D MHD simulation initially also has significant surface density at the 3:1 resonance $r = 15.2$, comparable to the higher alpha 2D simulations, since the manually added magnetic field loops act to increase angular momentum transport. However, the surface density at the resonance then drops and becomes more like the $\alpha=0.01$ 2D value, and the effective $\alpha$ due to magnetic stresses in the 3D simulation also declines to $\sim 0.01$. We can see the effect of this in the tidal contribution term in Figure~\ref{fig:mhd_3d_eccent_growth} (blue curve, bottom left). The tides initially try to increase eccentricity since there is enough mass near the 3:1 resonance, but as the disc becomes truncated, the tidal contribution diminishes and the blue curve flattens. In Figure~\ref{fig:dens_wave_compare}, the $(2, 3)$ density wave amplitude in the inner parts of the disc ($r \lesssim 12$) of the 3D simulation (purple curve) also decreases over time.

\begin{figure}
    \includegraphics[width=\columnwidth]{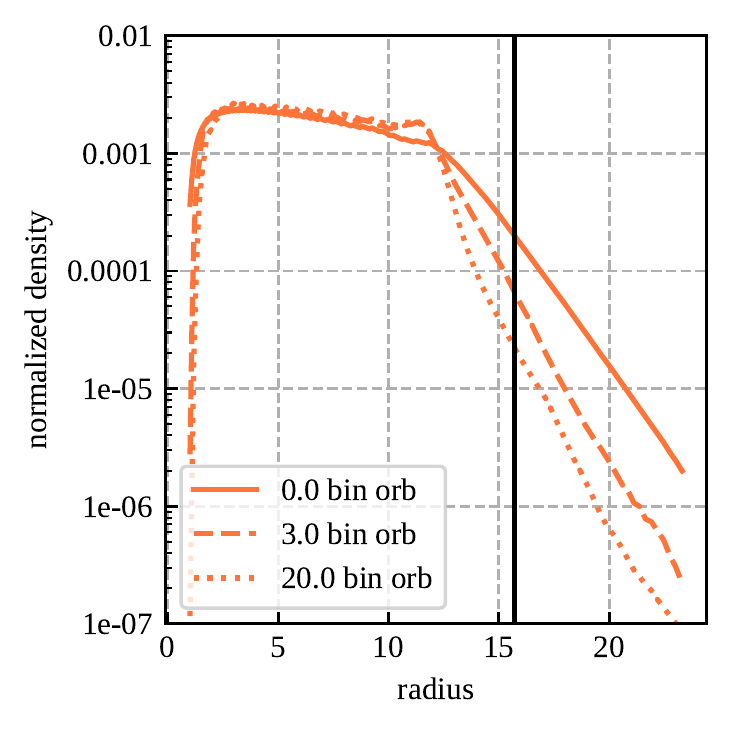}
    \caption{Surface density at different times of alpha0\_restart simulation. We see the surface density in outer regions of the disc rapidly decline as a result of the tidal truncation effect when viscosity isn't present to spread the disc. Vertical black line indicates the location of the nominal 3:1 resonance.}
    \label{fig:alpha0_restart_dens}
\end{figure}

\begin{figure}
    \includegraphics[width=\columnwidth]{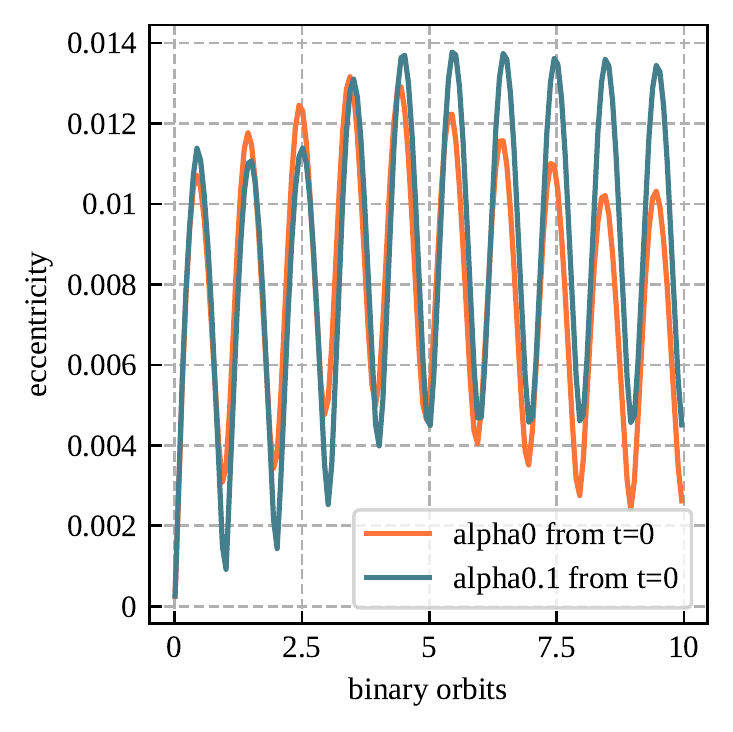}
    \caption{Eccentricity growth for alpha0\_restart simulation compared with initial evolution of alpha0.1 simulation. Eccentricity grows initially for the alpha0\_restart simulation but then declines as the disc becomes tidally truncated.}
    \label{fig:alpha0_restart_eccent_growth}
\end{figure}

We additionally ran a 2D simulation restarted from the $\alpha=0.1$ simulation data in order to investigate the tidal truncation phenomenon and its effect on eccentricity evolution. We axisymmetrized the surface density profile at $t=30$ binary orbits and used it to initialize the restarted simulation. Additionally we turned off the viscosity so $\alpha=0$ in the restarted simulation. The evolution of the surface density profile is plotted in Figure~\ref{fig:alpha0_restart_dens}. We see that without the large $\alpha$ viscosity keeping the disc spread, the surface density of the outer parts of the disc rapidly declines within a few binary orbits. The eccentricity evolution is also seen to be affected by the tidal truncation as seen in Figure~\ref{fig:alpha0_restart_eccent_growth}. Eccentricity initially grows for the restarted simulation (orange curve) since there is enough density in the outer parts of the disc, but as the disc becomes tidally truncated, eccentricity then declines.

\section{Discussion and conclusions}\label{sec:conclusion}
We ran a 3D MHD global simulation of an accretion disc modeled after a real AM CVn. One of the largest amplitude oscillations present in the real system's lightcurve is a superhump frequency, suggesting that it has an eccentric disc. However, after running our simulation for around 200 binary orbits, we see no evidence of significant eccentricity growth in our simulated disc. It should be cautioned however that our simulation has not run for a full viscous time and is likely at a much higher temperature compared to the real system. Running a global simulation of an AM CVn at a realistic temperature for a viscous time remains a significant computational challenge. \citet{lubow2010} used an eccentricity model to show that lowering the disk temperature resulted in faster eccentricity growth and also confined the eccentricity more in radius since the pressure forces compete with the resonance by spreading the local eccentricity over the disk.

Since earlier results of \citet{kley} found that in 2D simulations, increased viscosity and the absence of an accretion stream helps eccentricity to grow, we artificially introduced additional magnetic field loops and turned off the accretion stream in our simulation after $t=226$ binary orbits, but we still observed no significant eccentricity growth. To better understand this shortcoming of our 3D  MHD simulation, we ran three additional 2D simulations with artificial viscosities of $\alpha=0.01, 0.1, 0.2$ initialized using the surface density and velocity profiles of the 3D MHD simulation and compared the results against the 3D MHD simulation.

Of our three 2D simulations, only the $\alpha=0.1$ and $\alpha=0.2$ simulations showed significant eccentricity growth, whereas the $\alpha=0.01$ simulation did not, which is similar to earlier findings of \citet{kley} that found that a larger kinematic viscosity leads to more rapid eccentricity growth. To understand this dependence on viscosity better, we computed the direct contribution to eccentricity growth of each force present in our simulations. We found that the dominant driver for eccentricity growth comes from the companion's tidal field as expected. However, the viscosity in the two 2D simulations with $\alpha=0.1, 0.2$ also has a significant direct effect to grow eccentricity, especially in the $\alpha=0.2$ simulation where its contribution is roughly half of the tidal one. In contrast, however, in the 3D MHD simulation, the direct effect of the magnetic forces is actually to decrease eccentricity. Magnetic forces in MRI turbulence are usually thought of as the underlying physical mechanism behind the viscous spreading of the disc, which would aid the tidal growth of eccentricity. But here we find that their direct effect opposes eccentricity growth, opposite to the direct
action of artificial alpha viscosity in 2D simulations. Future MHD simulations are needed to determine whether this is a general phenomenon of MRI turbulence or specific to our setup.

Also noteworthy is that the $\alpha=0.2$ simulation has a more rapid apsidal precession rate compared to the $\alpha=0.1$ simulation, implicating a longer superhump period for the former. We computed the contribution of the viscous force on the apsidal precession rate and found it to be negligible. Instead the difference in precession rates comes from the tidal force, but since both simulations have the same tidal potential, it is likely that the cause of the difference is that the $\alpha=0.2$ simulation has more disc mass in the outer radii where the tidal effect is stronger. Hence, a higher alpha seems to result in a longer superhump period not because of the direct effect of the viscous force on precession but because of the larger disk resulting from the more efficient angular momentum transport.

In the 2D $\alpha=0.1$ simulation that showed significant eccentricity growth, we confirmed that eccentricity growth driven by the tidal potential occurs through the mode-coupling mechanism of \citet{lubow_theory}. From the simulation, we measured the direct contribution of each relevant spiral wave as they couple to the tidal potential to produce eccentricity. We find that the dominant contribution comes from the $(2, 3)$ wave of the form $e^{2 i \phi - 3 i \Omega_p t}$ excited by the 3:1 resonance, consistent with \citet{lubow_theory}.  However, we also find that several other waves also contribute a significant amount to the eccentricity evolution. The radial dependence of the $(2, 3)$ wave's contributions to eccentricity also shows oscillations beyond the eccentric Lindblad resonance in the evanescent region. In the 3D MHD simulation that does not show significant eccentricity growth, the $(2, 3)$ wave's amplitude is diminished in comparison to the 2D simulations that do have eccentricity growth.

Since these spiral waves are thought to be driven by the tidal potential, and the tidal potential is the same throughout all our simulations, the difference in wave amplitude is likely due to the different surface density distributions of our discs. We show that after an initial transient phase, the surface density at outer radii near the 3:1 resonance in the 3D MHD simulation is comparable to that of the $\alpha=0.01$ 2D simulation that did not show significant eccentricity growth, whereas the $\alpha=0.1, 0.2$ 2D simulations that had eccentricity growth had much more mass at outer radii. The effective $\alpha$ due to magnetic stresses in our 3D simulation also settles to $\sim 0.01$. Additionally, when we initialize a 2D simulation with the surface density of the $\alpha=0.1$ simulation but turn off viscosity, we see that though eccentricity grows initially, mass rapidly falls inward as the disc is tidally truncated, and eccentricity growth ceases.

Taken together, this could suggest that an effective alpha of $\alpha \sim 0.01$, commonly seen in MHD simulations without a net poloidal field, may not be high enough to spread sufficient mass to larger radii compared to real white dwarf accretors to enable eccentricity growth, or that the magnetic field is modeled incorrectly if the magnetic stresses are always completely suppressing eccentricity growth, though future simulations are needed to explore this further. This shortcoming of MHD simulations has been suggested previously in the context of dwarf novae \citep{KIN07,KOT12}, although convection may provide a resolution to this problem in that context \citep{HIR14,SCE18}. Effective alphas measured in observations of dwarf nova outbursts give an estimate of $\alpha \sim 0.1-0.2$. The low effective alpha seen in our MHD simulation seems to also be the cause for the failure
to produce the eccentricity responsible for one of the larger amplitude periodicities in the lightcurve of some of these accreting white dwarf systems. 
It is well known that MRI turbulence with net poloidal magnetic fields can produce an effective $\alpha$ that is much larger than $1\%$ \citep{HGB1995,Mishraetal2020}, which also depends on the amount of poloidal flux in the disk. It will be interesting to explore the eccentricity of the disk with poloidal magnetic fields, which can either come from the white dwarf or the companion, for future MHD simulations.

\section*{Acknowledgements}
The authors thank the anonymous referee for many helpful comments and suggestions for future work.

This work was supported by NASA Astrophysics Theory Program grant 80NSSC18K0727. Resources supporting this work were provided by the NASA High-End Computing (HEC) Program through the NASA Advanced Supercomputing (NAS) Division at Ames Research Center.

Use was also made of computational facilities purchased with funds from the National Science Foundation (CNS-1725797) and administered by the Center for Scientific Computing (CSC). The CSC is supported by the California NanoSystems Institute and the Materials Research Science and Engineering Center (MRSEC; NSF DMR 1720256) at UC Santa Barbara.

The Center for Computational Astrophysics at the Flatiron Institute is supported by the Simons Foundation.

\section*{Data Availability}
All our simulation data in the form of HDF5 files is available upon request.

\bibliographystyle{mnras}
\bibliography{ref}

\bsp
\label{lastpage}

\end{document}